\documentclass[9pt]{extarticle}

\usepackage{graphicx} 
\usepackage{amsmath}
\usepackage{adjustbox}
\usepackage{booktabs}
\usepackage{subcaption} 
\usepackage{authblk}
\usepackage{newunicodechar}
\newunicodechar{−}{\ensuremath{-}}

\usepackage{multicol}
\usepackage{float}
\usepackage[style=numeric-comp, sorting=none]{biblatex}
\usepackage{csquotes}

\usepackage[british]{babel}
\usepackage{multirow}
\usepackage{dblfloatfix}

\usepackage[letterpaper,top=2cm,bottom=2cm,left=3cm,right=3cm,marginparwidth=1.75cm]{geometry}

\usepackage{amssymb}
\usepackage[colorlinks=true, allcolors=blue]{hyperref}
\usepackage{setspace}

\bibliography{references}

\title{ \textbf{Odd Physics Off the Diagonal: Constraining CP-violating SMEFT with Quantum Tomography}}

\author[1]{Avalon Roberts\thanks{Contact: \href{mailto:avalon.roberts@postgrad.manchester.ac.uk}{avalon.roberts@postgrad.manchester.ac.uk}}}
\author[2]{Patrick Dougan\thanks{Contact: \href{mailto:douganp@smu.edu}{douganp@smu.edu}}}
\author[1]{Alexander Oh}
\author[1]{Savanna Shaw}
\affil[1]{Department of Physics \& Astronomy, University of Manchester, Manchester, M13 9PL, UK}
\affil[2]{Department of Physics, Southern Methodist University, Dallas, TX 75275-0175, USA}

\date{}

\makeatletter

\renewcommand{\abstract}{%
    \small\quotation\noindent
}

\makeatother

\begin{document}

\pagenumbering{arabic}
\setcounter{page}{1}

\maketitle
\vspace{-4em}
\begin{abstract}

New sources of charge-parity (CP) violation beyond those described in the Standard Model (SM) are required to explain the observed matter--antimatter asymmetry of the Universe. The Standard Model Effective Field Theory (SMEFT) provides a framework to introduce additional electroweak sources of CP-odd physics in a model-independent manner. However, these CP-violating signatures are mostly degenerate to CP-even SMEFT operators in polarisation-blind observables, distinguishable only in the SM-New Physics (NP) interference using the azimuthal decay angle. Using Quantum Tomography techniques, we present a new approach to constraining these NP effects. Reconstructing the spin density matrix (SDM) of a diboson system, we go beyond `interference resurrection' to exploit the full signature of the Beyond-SM physics, including the pure quadratic NP terms. We show that this approach provides superior simultaneous sensitivity to characteristic features of CP-even and CP-odd contributions, including effects not fully captured by traditional angular observables.
\end{abstract}

\begin{multicols}{2}

\section{Introduction}

The observed matter-antimatter asymmetry of the Universe remains an important unsolved problem of particle physics. A necessary condition for generating this asymmetry is the violation of charge-parity (CP) symmetry, as outlined by Sakharov in 1967~\cite{sakharov1967violation}. While some CP violation is generated in the Standard Model (SM) through quark mixing via the CKM matrix, it is insufficient to account for the scale of the baryon asymmetry of the Universe~\cite{gavela1994standard}. This shortfall motivates the search for new sources of CP violation beyond the SM.

Searches for CP violation in the weak gauge sector have been carried out at the LHC across a range of processes, including vector-boson fusion production of the Higgs boson~\cite{ATLAS:2016cpVBF, ATLAS:2020cpTau, CMS:2021cpTau} and diboson production~\cite{ATLAS:2020qdt}. \(WZ\) production is a particularly well-motivated channel for such searches, as it proceeds via the charged triple gauge vertex and is thus directly sensitive to CP-odd operators in the electroweak sector. Polarisation states and CP-sensitive observables in \(WZ\) production have been studied by ATLAS and CMS at the LHC~\cite{ATLAS:2019bsc, CMS:2021icx, ATLAS:2022oge}, with the first dedicated measurement of CP-violation sensitive observables in \(WZ\) reported recently~\cite{ATLAS:2025edf}. It should be noted that bounds on these CP-odd operators from the electron electric dipole moment (EDM) are currently much stronger than those set by LHC experiments, though LHC searches can contribute complimentary constraints to global fits~\cite{Azatov:2019xxn, Cirigliano:2019vfc}.

A systematic, model-independent framework for exploring such effects is provided by the Standard Model Effective Field Theory (SMEFT)~\cite{Isidori:2023pyp,Brivio2019}, which extends the Standard Model Lagrangian, $\mathcal{L}_{\text{SM}}$, with higher-dimensional operators constructed from SM fields respecting the SM gauge symmetries. The SMEFT Lagrangian can be written as
\begin{equation}
\mathcal{L}_{\text{SMEFT}} = \mathcal{L}_{\text{SM}} + \sum_{n=5}^{\infty} 
\frac{c_n}{\Lambda^{n-4}} \mathcal{O}_n,
\label{eq:smeft_lag}
\end{equation}
where $\mathcal{O}_n$ are operators of mass dimension $n$, $c_n$ are dimensionless Wilson coefficients encoding the effects of heavy new physics, and $\Lambda$ denotes the characteristic scale of new physics.

Assuming baryon and lepton number conservation, dimension-six operators provide the leading contributions from physics beyond the Standard Model. This assumption is well motivated by the absence of experimental evidence for baryon or lepton number violating processes, such as proton decay and neutrino-less double beta decay, which are constrained to occur only at very high energy scales, beyond those currently probed in proton-proton collisions at the LHC~\cite{SuperK:2016exg,KamLAND-Zen:2016pfg}. These operators can be organised into a complete and non-redundant set. A commonly adopted choice is the Warsaw basis~\cite{Grzadkowski2010}, which defines a minimal set of baryon-number-conserving operators invariant under the SM gauge symmetries. This basis is used throughout this study. In this case, the squared matrix element for a given process can be expanded as
\begin{equation}
|M|^2 = |M_{\text{SM}}|^2 + 2\,\mathrm{Re}\left\{ M_{\text{SM}} M_{D6}^* \right\} 
+ |M_{D6}|^2,
\label{eq:smeft_ME}
\end{equation}
where $M_{\text{SM}}$ is the SM matrix element, the interference term $\mathrm{Re}\{ M_{\text{SM}} M_{D6}^* \}$ encodes the leading linear sensitivity to new physics, and $|M_{D6}|^2$ represents quadratic contributions arising purely from dimension-six operators.

The set of CP-odd operators, $\widetilde{\mathcal{O}}_i$, is of particular interest, as it provides a parametrisation of potential new sources of CP violation. In recent years, significant effort has been devoted to developing methods that enhance sensitivity to such operators~\cite{PhysRevD.98.052005, ATLAS:2020qdt, ATLAS:2021wwjj, CMS:2021cpTau, ATLAS:2016cpVBF, CMS:2019HVV, ATLAS:2020cpTau}, including approaches based on machine learning~\cite{Bhardwaj:2021ujv,PhysRevD.107.016008,Brehmer:2018eca, Cranmer:2015bka,Cruz:2024grk}. A subset of these operators yields a complete parametrisation of CP violation in weak gauge-Higgs interactions and is closed under renormalization group evolution at dimension-six, ensuring a consistent framework for probing CP-violating effects in this sector at all energy scales~\cite{Grzadkowski2010,Alonso:2013hga}.
\begin{equation}
\begin{split}
\mathcal{O}_{\varphi \widetilde{W}} &= (\varphi^\dagger \varphi) 
    \widetilde{W}^{i,\mu\nu} W^i_{\mu\nu}, \\
\mathcal{O}_{\varphi \widetilde{B}} &= (\varphi^\dagger \varphi) 
    \widetilde{B}^{\mu\nu} B_{\mu\nu}, \\
\mathcal{O}_{\varphi \widetilde{W}B} &= (\varphi^\dagger \tau^I \varphi) 
    \widetilde{W}^{i,\mu\nu} B_{\mu\nu}, \\
\mathcal{O}_{\widetilde{W}} &= \epsilon_{ijk} 
    \widetilde{W}^i_{\mu\nu} W^{j,\nu\rho} W_\rho^{k,\mu}.
\end{split}
\label{eq:CPoddOps}
\end{equation}
The CP-even counterpart to this subset is the group of operators
\begin{equation}
\begin{split}
\mathcal{O}_{\varphi W} &= (\varphi^\dagger \varphi) W^{i,\mu\nu} W^i_{\mu\nu}, \\
\mathcal{O}_{\varphi B} &= (\varphi^\dagger \varphi) B^{\mu\nu} B_{\mu\nu}, \\
\mathcal{O}_{\varphi WB} &= (\varphi^\dagger \tau^I \varphi) W^{i,\mu\nu} B_{\mu\nu}, \\
\mathcal{O}_{W} &= \epsilon_{ijk} W^i_{\mu\nu} W^{j,\nu\rho} W_\rho^{k,\mu},
\end{split}
\label{eq:CPevenOps}
\end{equation}
where $\varphi$ denotes the Higgs doublet field, and $(\varphi^\dagger \varphi)$ is a gauge-invariant scalar under the electroweak symmetry group $SU(2)_L \times U(1)_Y$. The tensors $W^{i,\mu\nu}$ and $B^{\mu\nu}$ are the field strength tensors corresponding to the $SU(2)_L$ and $U(1)_Y$ gauge fields, respectively.

\begin{figure}[H]
  \centering
  \includegraphics[width=0.48\textwidth]{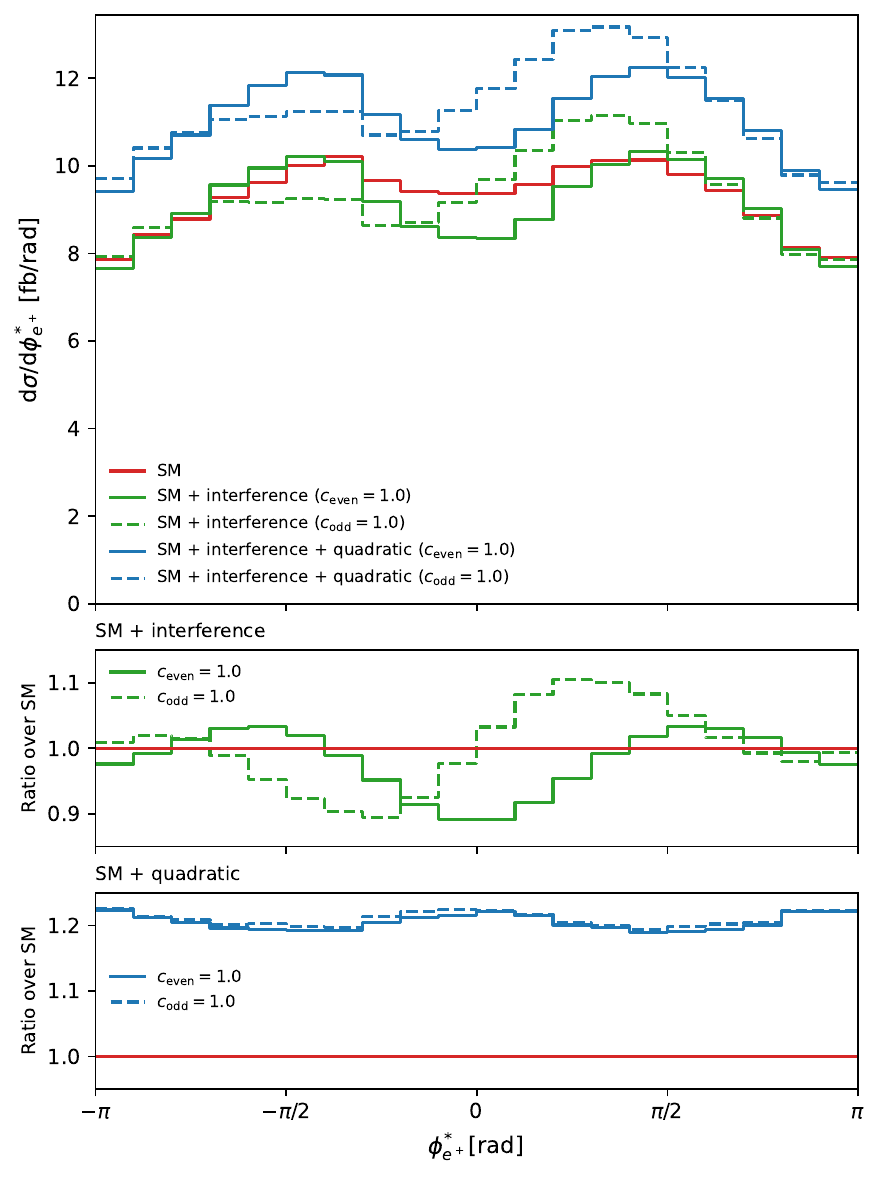}
  \caption{The $\phi_{truth}$ distribution at LO for the process 
  $pp \to W^+Z \to e^+ \nu_e \mu^+ \mu^-$ with the inclusive setup of Ref.~\cite{ElFaham:2024uop}, applying the invariant mass cut $81~\text{GeV} < M_{\mu^+\mu^-} < 101~\text{GeV}$. The upper panel shows the differential cross section $d\sigma/d\phi_e^*$ in units of fb, comparing the SM prediction (red) with the SM + interference and SM + interference + quadratic contributions for $c_{even} = 1.0$ (solid) and $c_{odd} = 1.0$ (dashed). The middle and lower panels show the ratio to the SM for the interference and quadratic-only contributions, respectively.}
  \label{fig:phi_neutrino_truth}
\end{figure}

In the limit of small Wilson coefficients, the leading BSM contribution in a SMEFT interpretation comes from the SM-BSM interference term, which scales as $\Lambda^{-2}$, while the purely quadratic BSM term, scaling as $\Lambda^{-4}$, is suppressed. However, it was shown in Ref.~\cite{Azatov_2017} that in the high-energy limit, this effect does not appear in polarisation-blind measurements for processes involving transversely polarised bosons. 

\begin{figure}[H]
    \centering
    \includegraphics[width=0.8\linewidth]{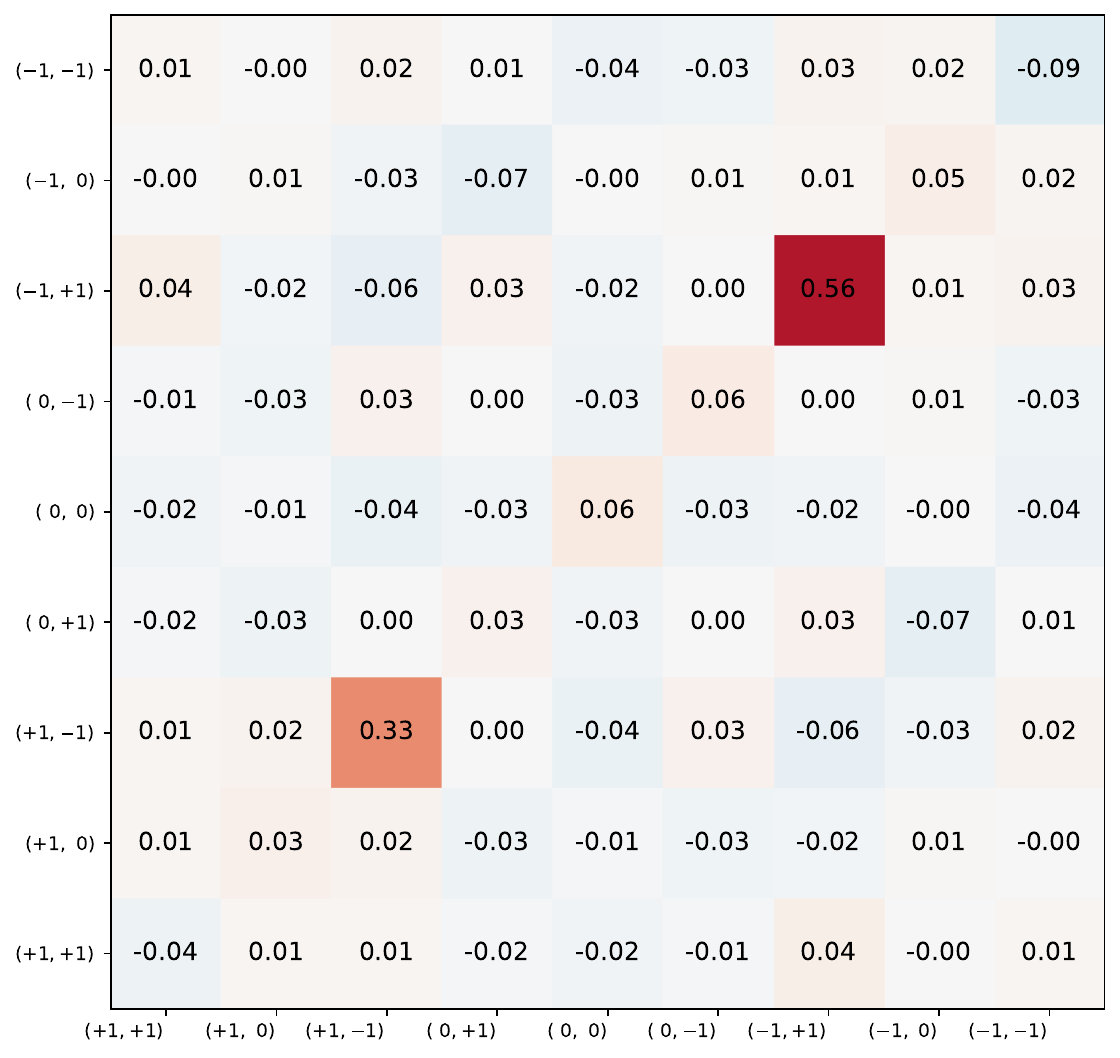}
    \caption{Real part of the spin density matrix for the \(W^+Z\) system in the SM, shown in the helicity basis.}
  \label{fig:sm_sdm}
\end{figure}

In the \(WZ\) processes studied in Ref.~\cite{Panico_2018} the vector bosons have final-state helicity ($\pm \mp$ ) in the SM and ($\pm \pm$) in the BSM physics. Anticipating our results, we see this effect in the comparison between figure \ref{fig:sm_sdm} and figures \ref{fig:sdm_quad_minus_even_real} and \ref{fig:sdm_quad_minus_odd_real}, where the difference in helicity structure can be seen by comparison of the diagonal elements of the density matrix. This helicity selection rule arises from the structure of the high-energy amplitude expansion~\cite{Shadmi:2018xan}. 

Following this result, it was demonstrated in Ref.~\cite{Panico_2018} that the interference term can be `resurrected' by selecting observables sensitive to the polarisation of the intermediate bosons. This resurrection can be seen through angular variables defined in the centre-of-mass frame of the diboson system, where the production angle and decay angles of the vector bosons allow for sensitivity to the interference contribution~\cite{Panico_2018,Banerjee:2019twi}. In particular, the azimuthal decay angle $\phi$ of a vector boson has been shown to be effective. This angle is defined in the rest frame of the decaying vector boson, as described in Ref.~\cite{Panico_2018}. However, $\phi$ provides limited separation between CP-even and CP-odd operators in the quadratic contribution $|M_{D6}|^2$. As shown in Figure~\ref{fig:phi_neutrino_truth}, the $\phi$ distributions for $O_W$ and $O_{\widetilde{W}}$ arising from the quadratic term have limited discrimination power.

A natural handle for separating the two classes of operators is provided by the structure of their amplitudes. Through the relation between amplitudes and their conjugates, Ref.~\cite{Panico_2018} shows that interference with CP-even BSM amplitudes results in a real contribution to the matrix element, while interference with CP-odd BSM amplitudes produces purely imaginary ones,
\begin{equation}
(\mathcal{A}_h)^* = \rho_{\textsubscript{CP}} \mathcal{A}_h,
\label{eq:SDMAmp}
\end{equation}
where \(\rho_{\textsubscript{CP}} = +1\) for CP-even amplitudes and \(\rho_{\textsubscript{CP}} = -1\) for CP-odd amplitudes. This structure is directly encoded in the off-diagonal elements of the spin density matrix, motivating the use of quantum tomography to construct a more powerful discriminating observable.

There has been much interest recently in applications of quantum tomography~\cite{DAriano_2003, Paris_Rehacek_2004, Lvovsky_Raymer_2009} and the use of quantum information techniques in particle physics~\cite{Ashby_Pickering_2023, Fabbrichesi:2021npl,Zhang:2025mmm} including as a probe of NP~\cite{Aoude:2023hxv,LoChiatto:2024dmx}, with entanglement between two top quarks having been demonstrated for the first time at the Large Hadron Collider in 2024~\cite{ATLAS2024entanglement}. Central to quantum tomography is the reconstruction of the spin density matrix (SDM), which encodes the full spin state of a system, from experimental data~\cite{DAriano_2003, Paris_Rehacek_2004}. By reconstructing the SDM of the diboson system, we go beyond a single angular projection to probe the complete spin structure of the final state, exploiting both the interference and quadratic contributions to gain discriminating power between CP-even and CP-odd operators where $\phi$ alone falls short.

Density matrices are often used to describe mixed spin states, where there is a degree of statistical uncertainty about the spin state of the system. By definition, the SDM $\rho$ can be expressed as a sum over outer products $\rho = \sum_i p_i |\psi_i\rangle \langle \psi_i|$, where $p_i$ are the probabilities associated with the system being in a spin state $|\psi_i\rangle$. The expectation value of an operator \( \hat{O} \) for a mixed state described by the density matrix \( \rho \) is then given by \( \langle \hat{O} \rangle = \text{Tr}(\rho \hat{O}) \)~\cite{NielsenChuang:2000}.

\section{Quantum Tomography of \texorpdfstring{\(WZ\)}{WZ} Production}

We apply quantum state tomography to the process $pp \to W^+Z \to e^+ \nu_e\mu^+ \mu^-$, probing the spin state of this system as a whole. The tomography formalism described by Ref.~\cite{Ashby_Pickering_2023} is used throughout and we restate the key results of that study here, including the 'Wigner P' functions, to aid the readability of this work.

The SDM was constructed under both SM and SMEFT hypotheses. We show that sensitivity to the operator $O_{\widetilde{W}}$ can be enhanced by taking this approach, with interference and quadratic terms being studied. It is useful to parametrise the SDM in an orthogonal basis so that it may be constructed from Monte Carlo simulated events. A common choice is the Generalized Bloch Vector parametrisation.

For a $d$-dimensional quantum system, any density matrix $\rho$ can be expressed in terms of a generalized Bloch vector. Specifically, $\rho$ can be written as
\begin{equation}
\rho = \frac{1}{d} I_d + \frac{1}{2} \sum_{i=1}^{d^2 - 1} r_i \lambda_i,
\label{eq:Rho}
\end{equation}
where $I_d$ is the $d \times d$ identity matrix, $\{ \lambda_i \}_{i=1}^{d^2 - 1}$ are a basis for the Lie algebra $SU(d)$ consisting of traceless Hermitian matrices (such as the Pauli matrices for d = 2 or generalized Gell-Mann matrices for d greater than 2), and $\vec{r} = (r_1, r_2, \ldots, r_{d^2 - 1})$ is a real vector in $\mathbb{R}^{d^2 - 1}$ known as the generalized Bloch vector. The $\lambda_i$ satisfy the orthonormality condition $\mathrm{Tr}(\lambda_i \lambda_j) = 2 \delta_{ij}$ and span the space of all traceless Hermitian operators. Hence all density matrices can be constructed from these plus the identity. 

\subsection{The W Boson}

The SDM for a single spin-1 (d = 3) particle can be expanded in terms of the identity matrix, the eight Gell-Mann matrices $\{\lambda_i\}_{i=1}^8$, the generators of the $\text{SU}(3)$ Lie algebra
\begin{equation}
\rho = \frac{1}{3} I + \frac{1}{2} \sum_{i=1}^{8} a_i \lambda_i,
\label{eq:W_Rho}
\end{equation}
where $I$ is the $3 \times 3$ identity matrix, and the Fano coefficients $a_i$ form the components of the generalized Bloch vector $\vec{a} \in \mathbb{R}^8$. 

Due to the chiral nature of the interactions of the $W$ boson, its spin state can be inferred from the kinematic properties of its decays. As the $W$ boson couples only to leptons with left-handed chirality and anti-leptons with righted-handed chirality, its spin can be approximated as being in the direction of the lepton's momentum. This approximation is valid in the limit where the mass of the lepton is negligible compared with the mass of the $W$ boson $m_\ell \ll m_W$~\cite{Ashby_Pickering_2023}.

To extract $\vec{a}$ from kinematic observables, an inverted Wigner-Weyl formalism is used~\cite{Ashby_Pickering_2023}. This formalism provides a mapping from angular observables to the parameters of the spin density matrix. 

The `Wigner P' symbols of the Gell-Mann matrices are used here to map from the $W$ boson decay angles to the parameters of the SDM. Functions of the opening and azimuthal angles of the final-state leptons, $\theta$ and $\phi$, as defined in Ref.~\cite{Panico_2018}, the Wigner Ps are simply stated here, with their derivation from the application of the projection operator to an axis-aligned spin state given in Ref.~\cite{Ashby_Pickering_2023}: 
\begin{equation}
\begin{array}{l}
\Phi_1^{P\pm} = \sqrt{2}(5 \cos \theta \pm 1) \sin \theta \cos \phi \\
\Phi_2^{P\pm} = \sqrt{2}(5 \cos \theta \pm 1) \sin \theta \sin \phi \\
\Phi_3^{P\pm} = \frac{1}{4}(\pm 4 \cos \theta + 15 \cos 2\theta + 5) \\
\Phi_4^{P\pm} = 5 \sin^2 \theta \cos 2\phi \\
\Phi_5^{P\pm} = 5 \sin^2 \theta \sin 2\phi \\
\Phi_6^{P\pm} = \sqrt{2}(\pm 1 - 5 \cos \theta) \sin \theta \cos \phi \\
\Phi_7^{P\pm} = \sqrt{2}(\pm 1 - 5 \cos \theta) \sin \theta \sin \phi \\
\Phi_8^{P\pm} = \frac{1}{4\sqrt{3}} (\pm 12 \cos \theta - 15 \cos 2\theta - 5).
\end{array}
\label{eq:WignerPs}
\end{equation}
Here $\pm$ refers to the charge of the lepton produced by the W boson decay. The angles $\phi$ and $\theta$ are the decay angles of the vector boson in its rest frame. These angles are shown in Ref.~\cite{Panico_2018}. The parameters of the Bloch vector, or Fano coefficients, are then averages of these eight functions \cite{Ashby_Pickering_2023}
\begin{equation}
a_i = \frac{1}{2} \left\langle \Phi_i^{P\pm} \right\rangle_{\text{}}
\label{eq:WPAverage}
\end{equation}
where $i$ runs from one to eight. Using this formalism, the single boson spin density matrix can be fully reconstructed from experimental data, up to the ambiguity associated with reconstructing the longitudinal momentum component of the neutrino. 

\subsection{The Z Boson}

For the process $Z \to \ell^+ \ell^-$, the spin state of the $Z$ boson is partially encoded in the angular distribution of the final-state leptons. This is due to the chiral nature of the weak interaction, with the $Z$ boson coupling to both left- and right-handed leptons with couplings $c_L = -0.273$ and $c_R = +0.233$ ~\cite{Ashby_Pickering_2023}. 

Following the method presented in Ref.~\cite{Ashby_Pickering_2023}, the decay can be treated as a partial spin measurement. The angular observables can be related to the spin density matrix via generalized Wigner P-symbols, which are linear combinations of the standard symbols:
\begin{equation}
\Phi_{F_{Z \to \ell\ell},j}^P = \frac{1}{|c_R|^2 - |c_L|^2} \sum_k A_{jk} \Phi_k^{P\pm}.
\label{eq:generalized_p_z}
\end{equation}

The transformation matrix $A$ is given in Appendix~\ref{sec:a_matrix}. These generalised symbols can then be averaged over decay angles to extract the Bloch vector components and reconstruct the SDM of the $Z$ boson.

\subsection{The Diboson System}

The bipartite SDM, describing the joint spin state of two bosons can be expressed using the following Bloch vector parameterisation~\cite{Ashby_Pickering_2023}.
\begin{equation}
\begin{split}
\rho^{(d)} = \frac{1}{d} I_d &+
\sum_{i=1}^{d_1^2 - 1} a_i \lambda_i^{(d_1)} \otimes \frac{1}{d_2} I_{d_2}\\ 
&+ \sum_{j=1}^{d_2^2 - 1} \frac{1}{d_1} I_{d_1} \otimes b_j \lambda_j^{(d_2)} \\
&+ \sum_{i=1}^{d_1^2 - 1} \sum_{j=1}^{d_2^2 - 1} c_{ij} \lambda_i^{(d_1)} \otimes \lambda_j^{(d_2)},
\end{split}
\label{eq:DB_Rho}
\end{equation}
where $d_1$ and $d_2$ are the dimensions of the spin state spaces of the two bosons. In the case of two vector boson $d_1$ = $d_2$ = 3, and $d$ = $d_1$ x $d_2$ = 9, corresponding to the total spin degrees of freedom of the diboson system. The vectors $a_i$ and $b_i$ are Bloch vectors of the two bosons and can be obtained from the eight Wigner P symbols as described above. The matrix $c_{ij}$ describes correlations between the Bloch vectors of the bosons and can be calculated from products of the Wigner P symbols corresponding to each boson as follows:
\begin{equation}
{c}_{ij} = \left( \frac{1}{2} \right)^2 
\left\langle 
\widetilde{\Phi}^{P}_{F_l^{(A)}, i}(\hat{\mathbf{n}}_1) \,
\widetilde{\Phi}^{P}_{F_m^{(B)}, j}(\hat{\mathbf{n}}_2) 
\right\rangle_{\mathrm{av}}
\label{eq:c_ij}
\end{equation}

The quantities $\widetilde{\Phi}^{P}_{F_l^{(A)}, i}(\hat{\mathbf{n}}_1)$ and $\widetilde{\Phi}^{P}_{F_m^{(B)}, j}(\hat{\mathbf{n}}_2)$ are the Wigner P-symbols, evaluated for the angular configurations of bosons $A$ and $B$ respectively. The index $i$ labels the Gell-Mann matrices, while $F_l^{(A)}$ and $F_m^{(B)}$ denote the bosons.

\section{Monte Carlo Studies}

The process $pp \to W^+Z \to e^+ \nu_e \mu^+ \mu^-.$ was generated using the \texttt{MadGraph5\_aMC@NLO} framework~\cite{Alwall:2014hca} at leading-order (LO), which retains full spin-correlation information. This process was studied in order to access the usefulness of the methodology presented here. Other channels should be studied in future work. Next-to-leading-order (NLO) corrections are known to be large and real for \(WZ\) production, with their effects on the observables studied here well documented in Ref.~\cite{ElFaham:2024uop} and Ref.~\cite{Haisch:2025jqr}. The affects of these corrections on the the SDM has also been shown in the literature for several electroweak processes \cite{Pelliccioli:2026ltl,Grossi:2024jae,DelGratta:2025qyp,Goncalves:2025xer,Goncalves:2025mvl,DelGratta:2025xjp}. Future work should account for the inclusion of these effects and their impact on the usefulness of the SDM in constraining NP.

Two event selection schemes were considered, following the generation setup defined in Ref.~\cite{ElFaham:2024uop}. The inclusive setup applies a single cut on the invariant mass of the same-flavour opposite-sign lepton pair,
\begin{equation}
81~\text{GeV} < M_{\mu^+\mu^-} < 101~\text{GeV},
\label{eq:inclusive}
\end{equation}
which selects the $Z$ decay product while suppressing off-shell contributions. The fiducial setup imposes additional requirements designed to mimic realistic detector acceptance,
\begin{align}
&81~\text{GeV} < M_{\mu^+\mu^-} < 101~\text{GeV}, \nonumber\\
&p_{T,e^+} > 20~\text{GeV}, \quad p_{T,\mu^\pm} > 15~\text{GeV}, \nonumber\\
&M_{T,W} > 30~\text{GeV}, \quad |y_\ell| < 2.5, \nonumber\\
&\Delta R_{\mu^+\mu^-} > 0.2, \quad \Delta R_{\mu^\pm e^+} > 0.3,
\label{eq:fiducial}
\end{align}
where the transverse mass of the $W$ boson is defined as
\begin{equation}
M_{T,W} = \sqrt{2\, p_{T,e^+}\, p_{T,\text{miss}}
\left(1 - \cos\Delta\phi_{e^+,\text{miss}}\right)},
\label{eq:MTW}
\end{equation}
and $\Delta\phi_{e^+,\text{miss}}$ is the azimuthal separation between the positron and the missing transverse momentum. Results are presented for both setups; the fiducial case additionally requires neutrino reconstruction via the on-shell $W$-mass constraint. The NNPDF3.1 parton distribution functions~\cite{NNPDF3.1} were used to model the momentum distributions of the colliding protons in all event generation.
SM events were generated at a centre of mass energy of 13 TeV, yielding a cross section of $58.44 \pm 0.01\,\text{fb}$ in the fiducial setup described above. Reconstructed Gell-Mann parameters are given in the Appendix and are in good agreement with Ref.~\cite{barker2023quantum}.
The spin density matrix was constructed from these events. 

Small negative values along the central diagonal, which are unphysical, arise from this process being simulated only at leading order in QCD in this study. The SDM was constructed using SM events at next to leading order in QCD and these values were seen to become non-negative. 

\begin{figure*}[t]
    \centering
    \begin{subfigure}{0.49\textwidth}
        \centering
        \includegraphics[width=0.85\linewidth]{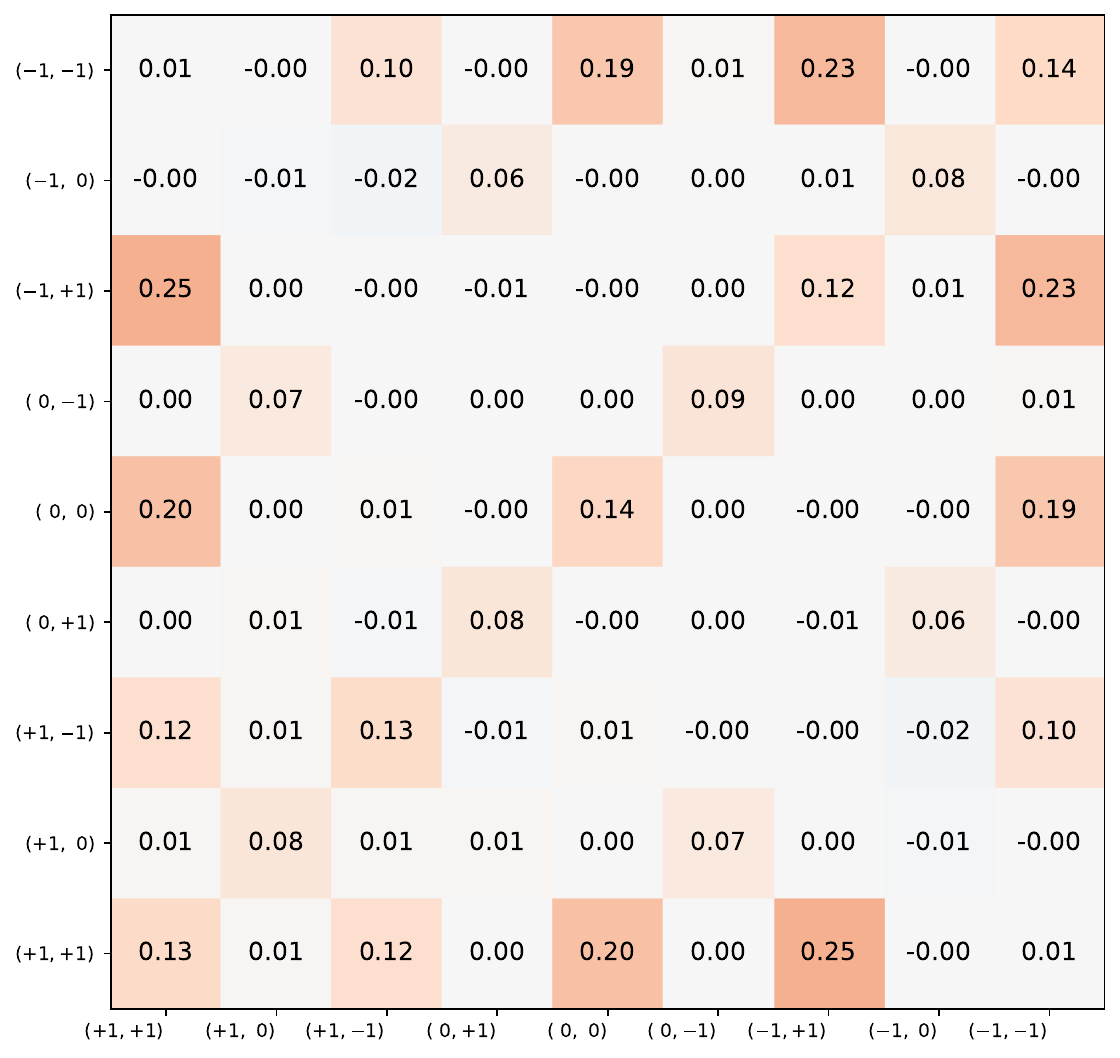}
        \caption{CP-even $O_W$: Real Component}
        \label{fig:minus_even_real}
    \end{subfigure}
    \hfill
    \begin{subfigure}{0.49\textwidth}
        \centering
        \includegraphics[width=0.85\linewidth]{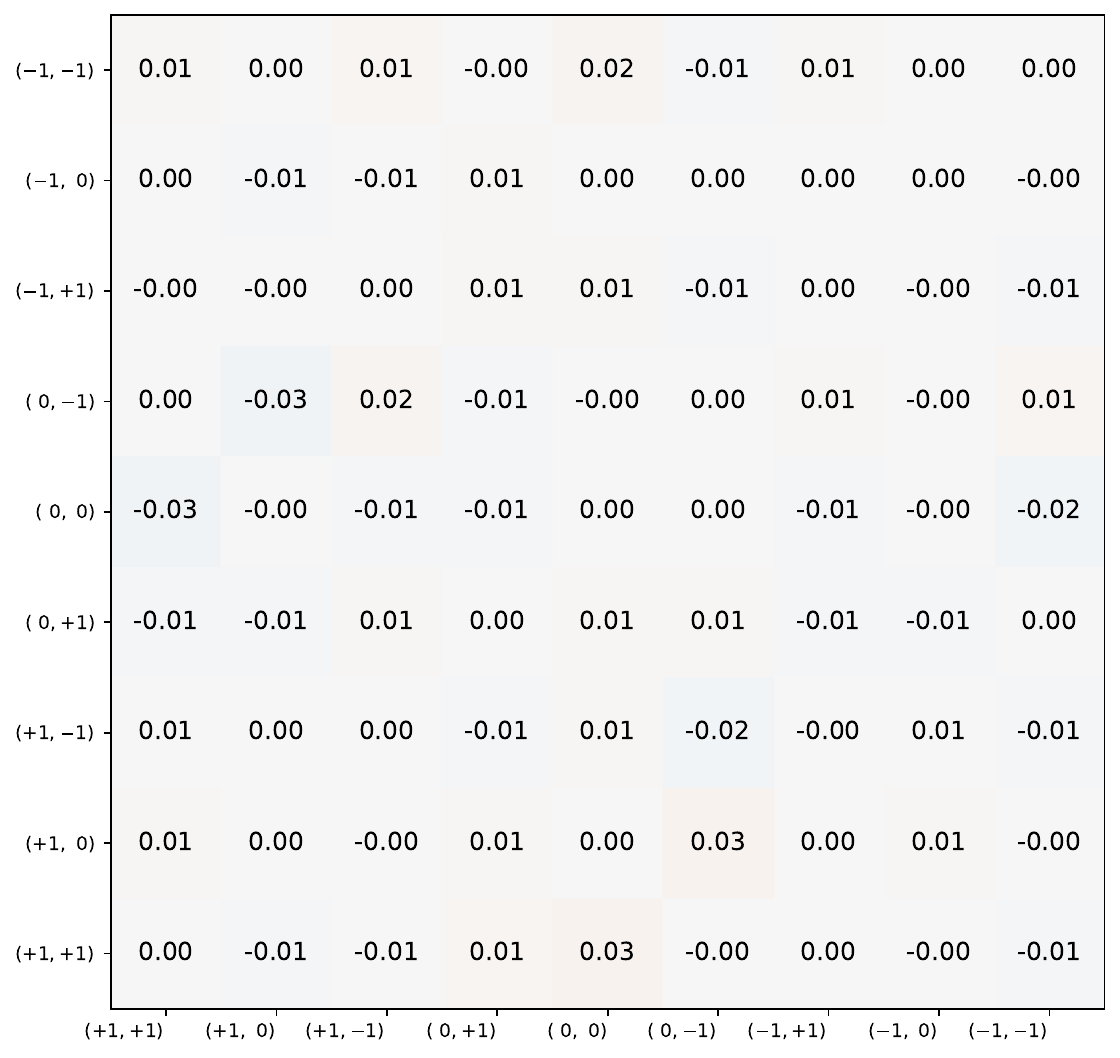}
        \caption{CP-even $O_W$: Imaginary Component}
        \label{fig:minus_even_imag}
    \end{subfigure}
    \vspace{1.5em}
    \begin{subfigure}{0.49\textwidth}
        \centering
        \includegraphics[width=0.85\linewidth]{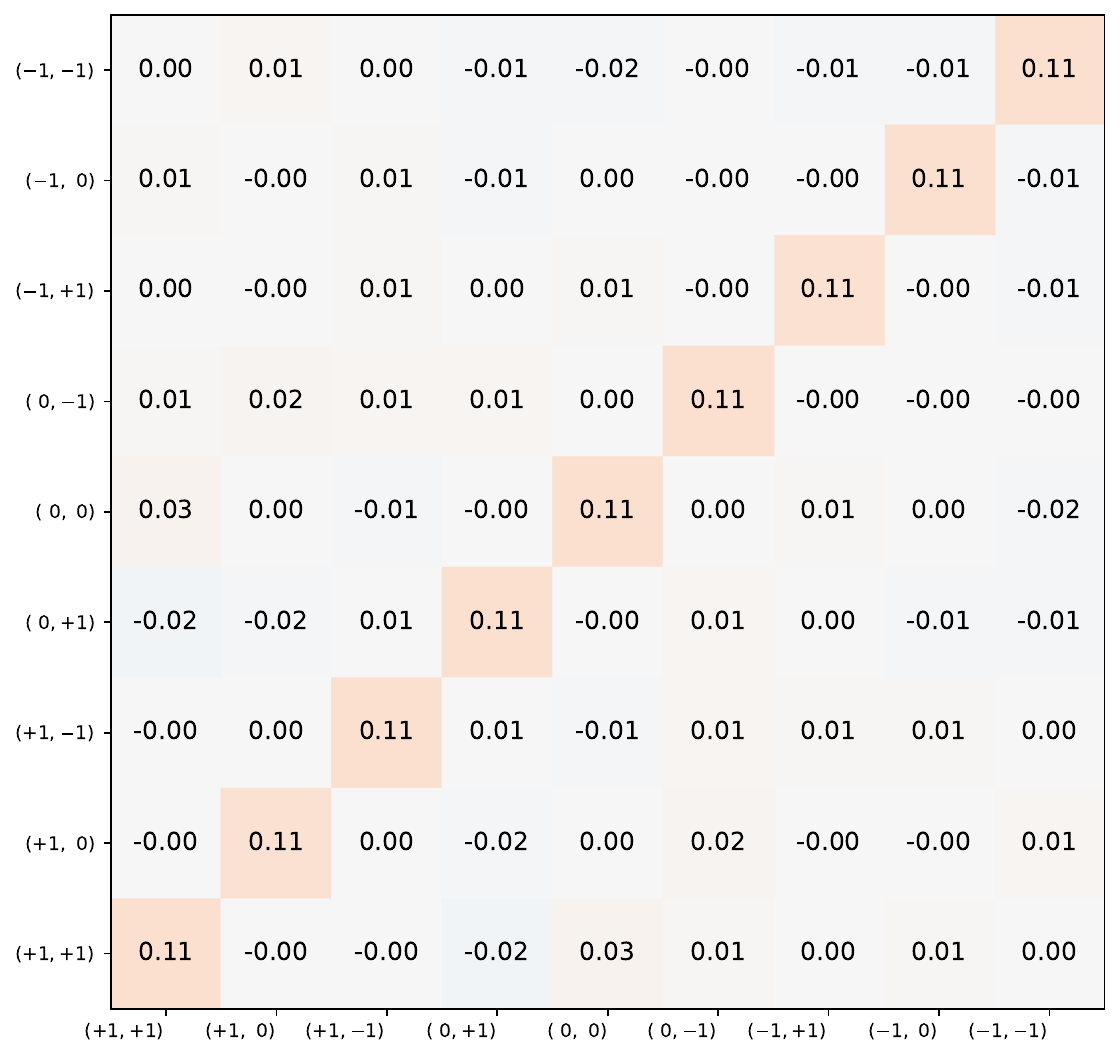}
        \caption{CP-odd $O_{\widetilde{W}}$: Real Component}
        \label{fig:minus_odd_real}
    \end{subfigure}
    \hfill
    \begin{subfigure}{0.49\textwidth}
        \centering
        \includegraphics[width=0.85\linewidth]{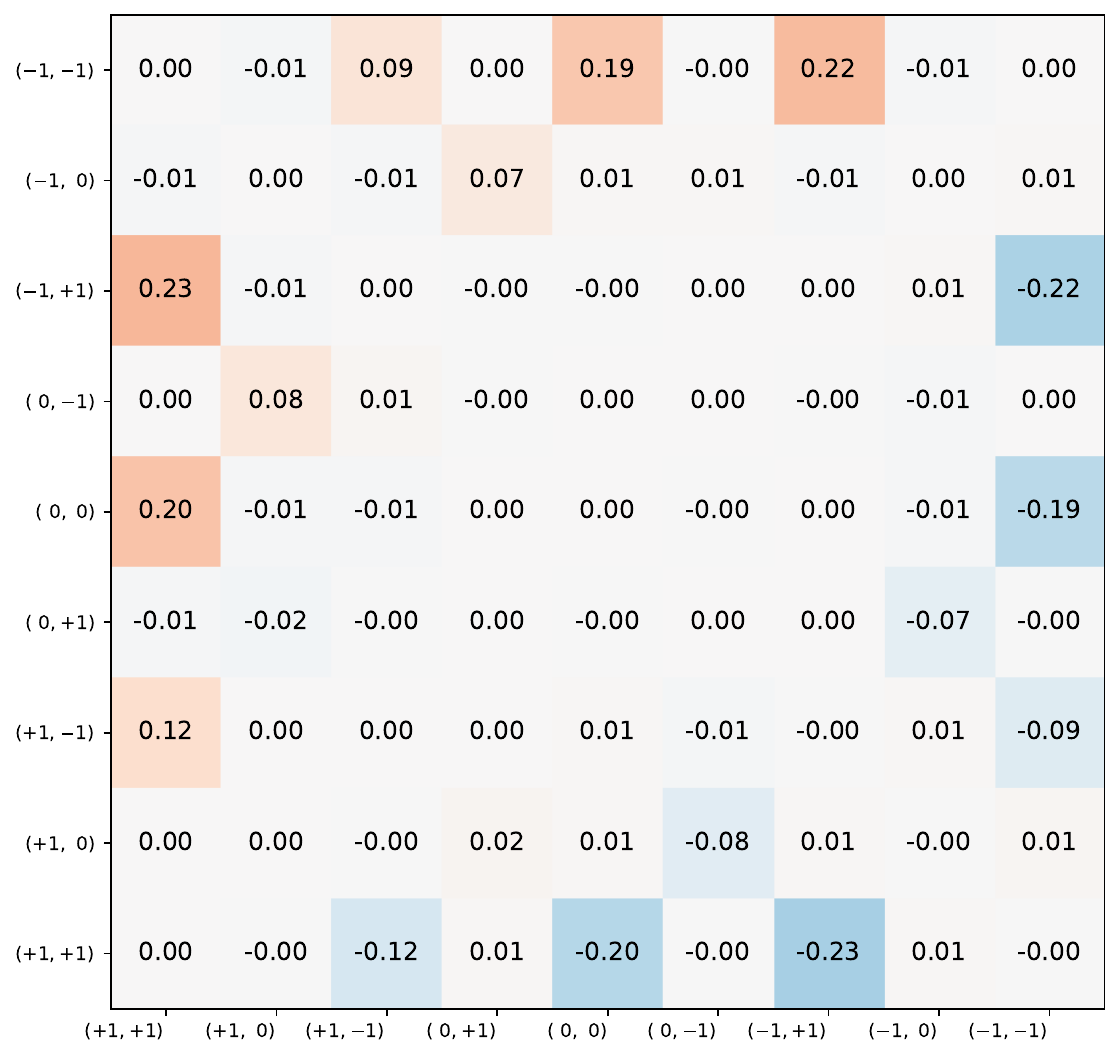}
        \caption{CP-odd $O_{\widetilde{W}}$: Imaginary Component}
        \label{fig:minus_odd_imag}
    \end{subfigure}
    \caption{The reconstructed SDMs for the interference terms of the CP-even $O_W$ (above) and CP-odd $O_{\widetilde{W}}$ (below) operators with the SM, separated into their real (left) and imaginary (right) components. The separation of the off-diagonal structure between the CP-odd and CP-even into the imaginary and real parts of the SDM means that this can be used to set 2D limits on them.}
    \label{fig:int_term}
\end{figure*}

Background processes and systematic uncertainties are not included in this study. These are neglected as we aim to demonstrate the sensitivity of the SDM to EFT effects in the \(WZ\) signal process itself; a full experimental analysis incorporating backgrounds such as \(ZZ\), \(t\bar{t}\), \(VVV\), and \(Z\)+jets production, as well as detector and theoretical systematic uncertainties, is left to future experimental work. This approach is consistent with other recent phenomenological studies of this kind~\cite{ElFaham:2024uop}, where fully-leptonic \(WZ\) is shown to yield high purity after the applied fiducial selection used in the recent ATLAS analyses~\cite{ATLAS:2019bsc, ATLAS:2022oge}, and where the primary conclusions rest on the shapes of kinematic distributions rather than on absolute event yields. The EFT limits we extract later should therefore be interpreted as indicative of the SDMs revelative discriminating power to other studied observables under idealised conditions.

SMEFT events were generated using the \texttt{SMEFTsim\_U35\_MwScheme\_UFO} model~\cite{Brivio2019, Brivio:2020onw} and the same SDM construction process was applied. As the events generated for the interference and quadratic terms are modifications to the SM, the SDMs constructed are not physical on their own. These should be considered additive modifications to the SM SDM, with the magnitude of the modification depending on the value of the Wilson coefficient. We obtained the following cross sections.
\begin{table}[H]
    \centering
    \begin{tabular}{lc}
    \hline
    Contribution & Cross section [fb] \\
    \hline
    SM & $58.44 \pm 0.01$ \\
    CP-even interference & $-1.07 \pm 0.01$ \\
    CP-even quadratic & $11.95 \pm 0.01$ \\
    CP-odd interference & $0.15 \pm 0.01$ \\
    CP-odd quadratic & $12.24 \pm 0.01$ \\
    \hline
    \end{tabular}
    \caption{Fiducial cross sections.}
    \label{tab:inclusive_xs}
\end{table}

\begin{figure*}[t]
    \centering
    \begin{subfigure}{0.49\textwidth}
        \centering
        \includegraphics[width=0.85\linewidth]{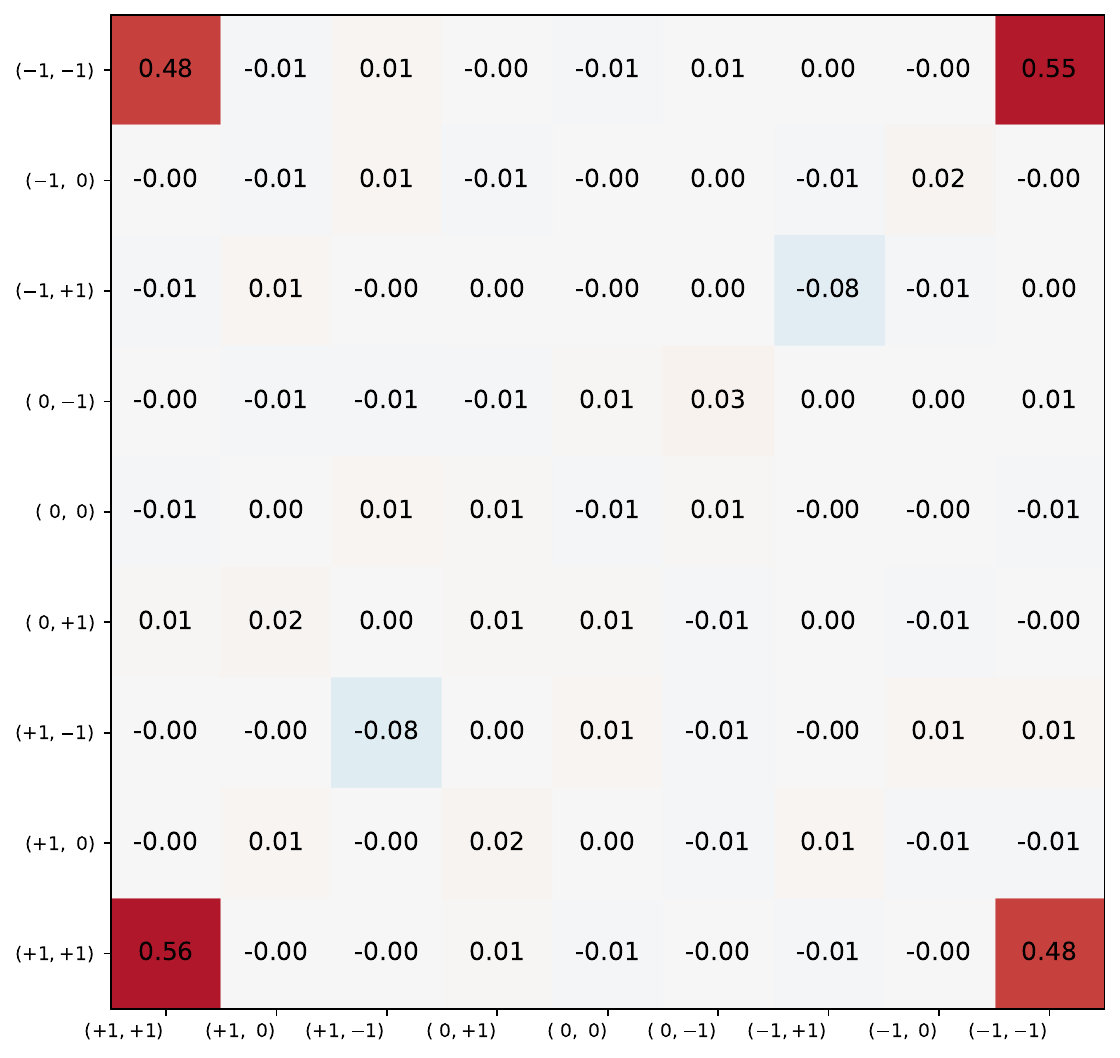}
        \caption{CP-even $O_W$: quadratic term, real component}
        \label{fig:sdm_quad_minus_even_real}
    \end{subfigure}
    \hfill
    \begin{subfigure}{0.49\textwidth}
        \centering
        \includegraphics[width=0.85\linewidth]{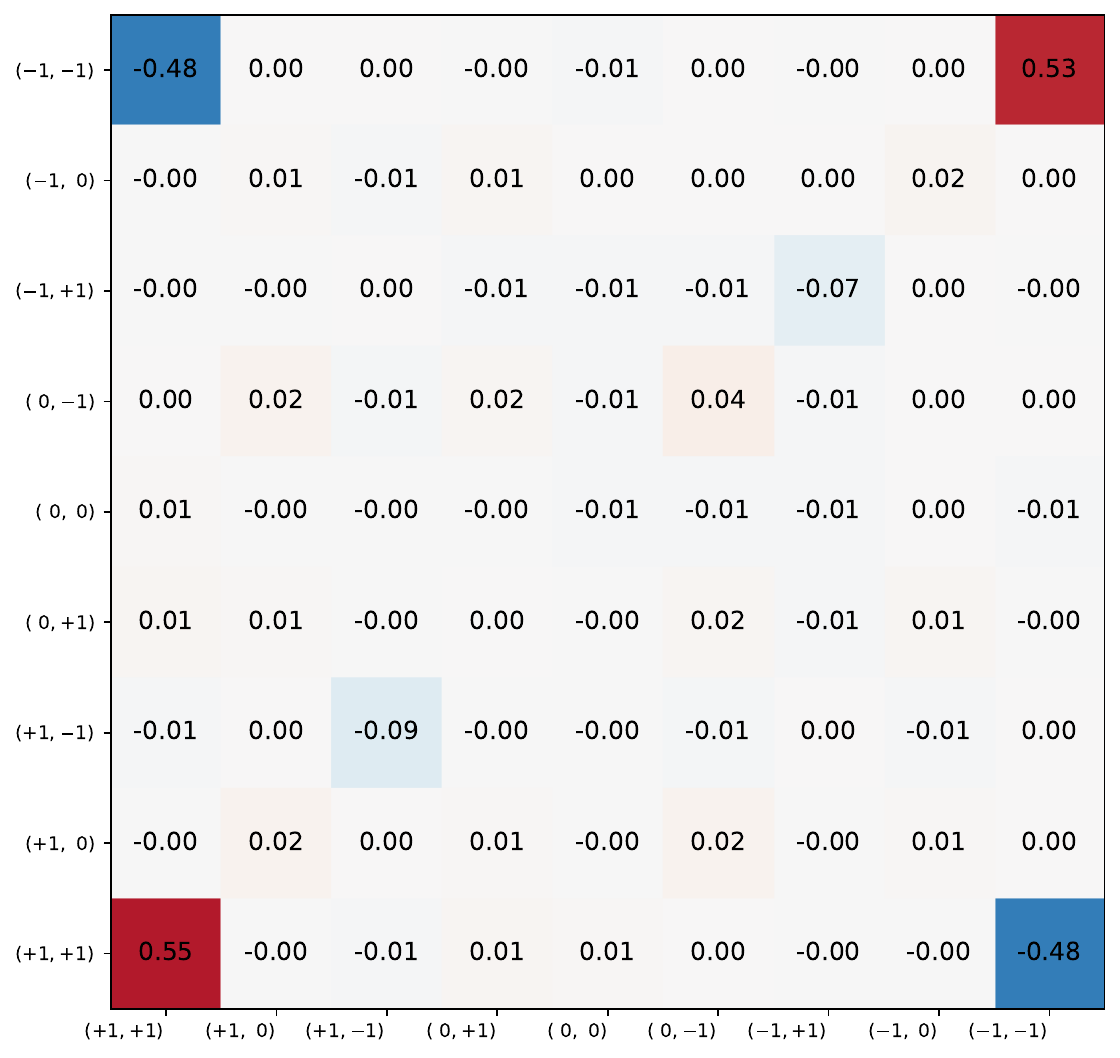}
        \caption{CP-odd $O_{\widetilde{W}}$: quadratic term, real component}
        \label{fig:sdm_quad_minus_odd_real}
    \end{subfigure}
    \vspace{1.5em}
    \begin{subfigure}{0.49\textwidth}
        \centering
        \includegraphics[width=0.85\linewidth]{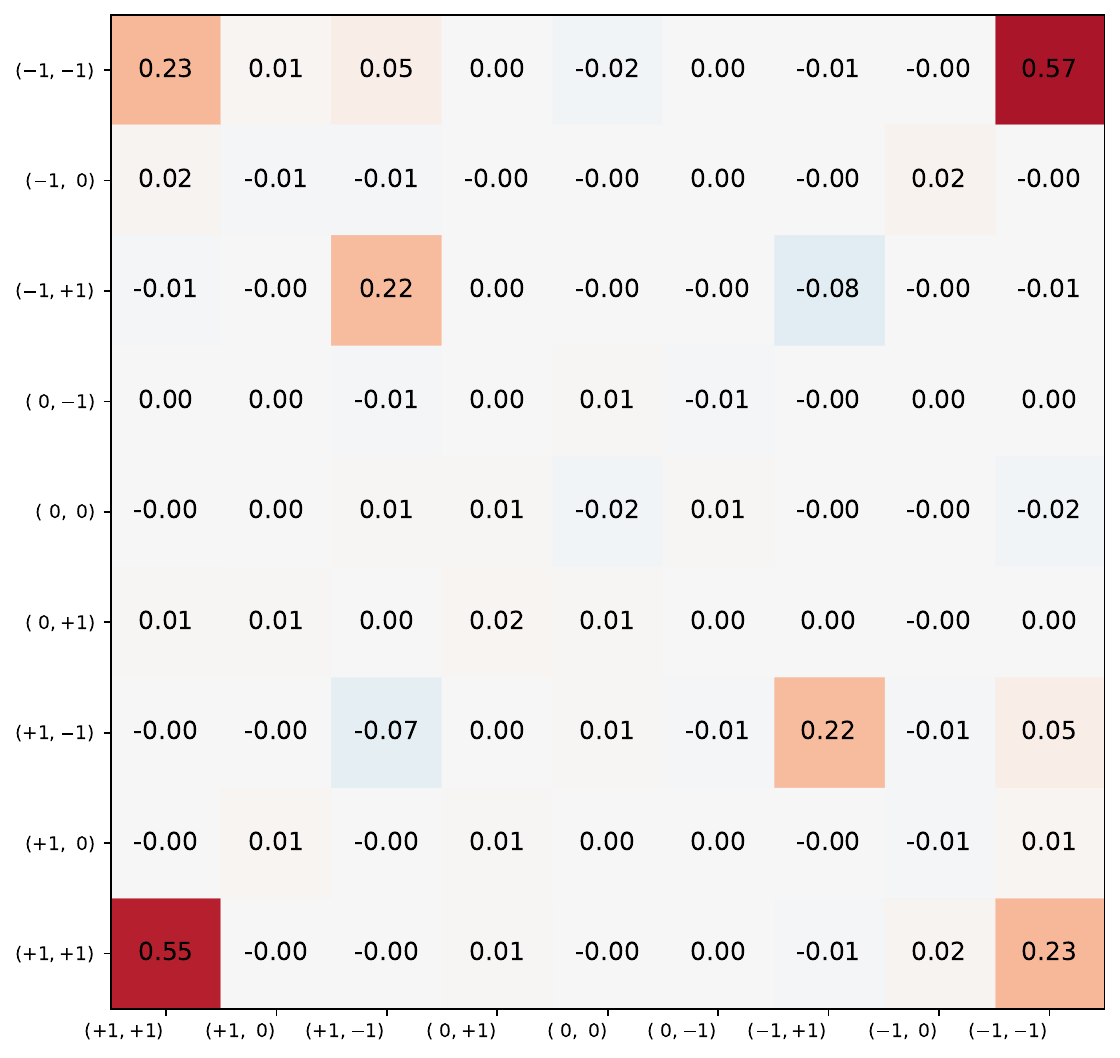}
        \caption{CP-even $O_{W}$: quadratic term with neutrino reconstruction, real component}
        \label{fig:cp_even_reco}
    \end{subfigure}
    \hfill
    \begin{subfigure}{0.49\textwidth}
        \centering
        \includegraphics[width=0.85\linewidth]{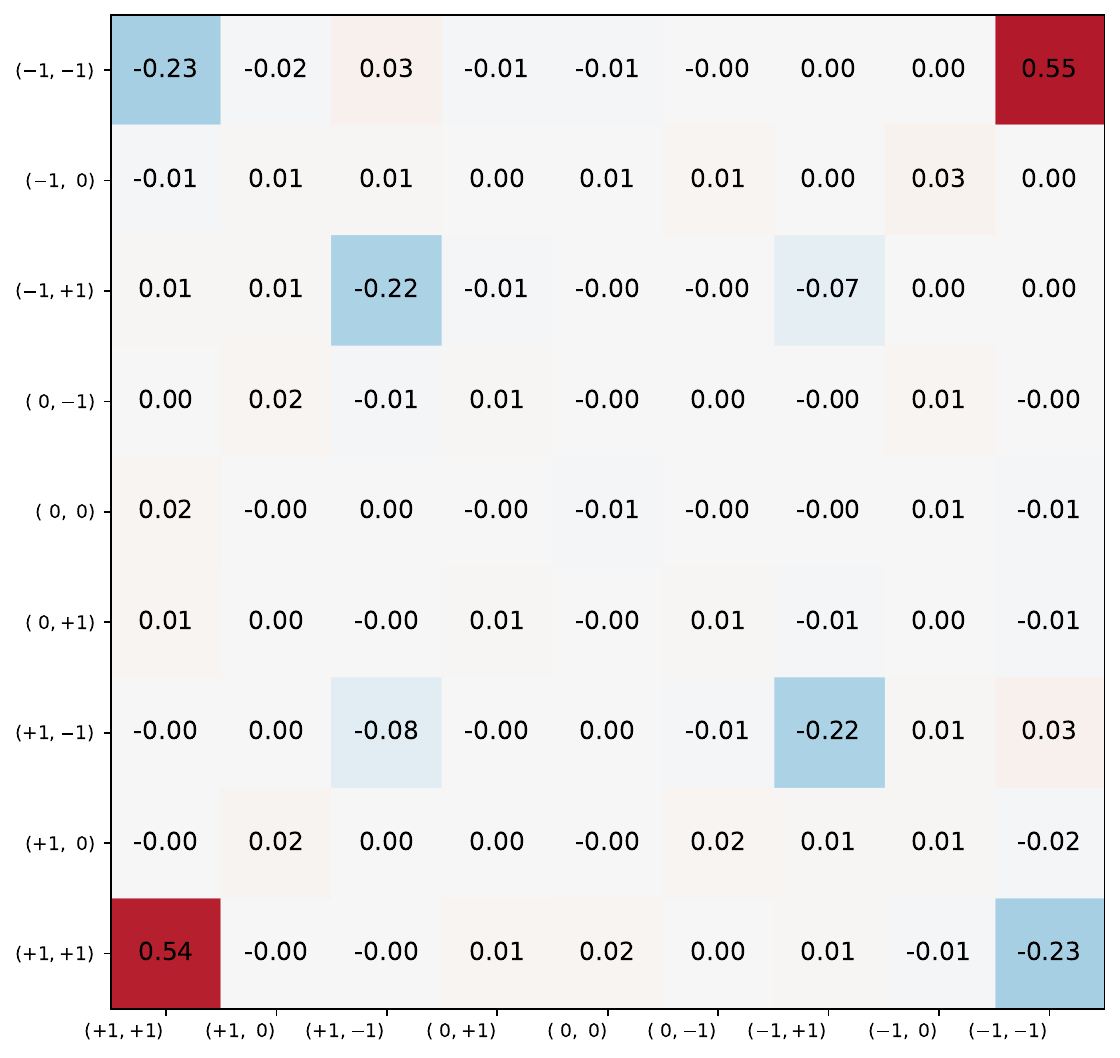}
        \caption{CP-odd $O_{\widetilde{W}}$: quadratic term with neutrino reconstruction, real component}
        \label{fig:cp_odd_reco}
    \end{subfigure}
    \caption{The reconstructed SDMs for the quadratic terms of the CP-even $O_W$ operator (left) and the CP-odd $O_{\widetilde{W}}$ operator (right) in the inclusive (top) selection setup and the fiducial (bottom) setup. It can be seen that some entries become split between matrix elements due to the ambiguity in the neutrino reconstruction. }
    \label{fig:quadratic_term}
\end{figure*}

The reconstructed SDM for the interference contribution is shown in Figures~\ref{fig:minus_even_real} \& \ref{fig:minus_even_imag} for $O_W$ and Figures~\ref{fig:minus_odd_real} \& \ref{fig:minus_odd_imag} for $O_{\widetilde{W}}$. Figure \ref{fig:wz_cij_sm} in the Appendix shows the reconstructed Gell-Mann parameters for this process in the SM, in good agreement with Ref.~\cite{barker2023quantum}. Figures \ref{fig:wz_cij_even_int}, \ref{fig:wz_cij_even_quad}, \ref{fig:wz_cij_odd_int}, and \ref{fig:wz_cij_odd_quad} show the Gell-Mann parameters for each component of the BSM physics.

These results are also in agreement with a result found in `Diboson Interference Resurrection' \cite{Panico_2018}. In the case of interference between the CP even operator, it can be seen that all entries in the matrix, i.e. products of amplitudes, are entirely real within the uncertainty bounds. In the case of interference between the SM and the CP odd operator, the diagonal elements of the matrix are real and non-zero.

The purely quadratic BSM contribution enhances deviations at high energy. Figures~\ref{fig:sdm_quad_minus_even_real} and ~\ref{fig:sdm_quad_minus_odd_real} show the reconstructed SDM for these events. The imaginary components are all zero within the statistical uncertainty, and therefore are not shown here for brevity. 

The operators $O_{\widetilde{W}}$ and $O_W$ clearly have different signatures in the off-diagonal elements of the quadratic spin density matrix whereas these operators have no discrimination power in angular decay distributions in the quadratic term~\cite{ElFaham:2024uop}. This motivated the construction of an observable, based on the spin density matrix, to optimally separate these contributions while retaining sensitivity to the cross-section enhancement from the quadratic component of the SMEFT operators.

\section{Neutrino Reconstruction}

To estimate the sensitivity to new physics that could be achieved in real data, we studied the effects on these new variables assuming the neutrino cannot be detected. Given that there is no initial transverse momentum in a collider environment, the transverse momentum of the neutrino can be inferred from kinematic constraints. This leaves only the longitudinal component unknown. 

In this study, we follow the neutrino reconstruction procedure of Ref.~\cite{Panico_2018}, whereby the rapidity of the neutrino can be constrained, up to a two-fold ambiguity, by making the assumption that the neutrino and the electron together exactly reconstruct the $W$ boson mass, i.e. that the $W$ boson is on-shell. 
The pseudo-rapidity of the neutrino can then be constrained as follows
\begin{equation}
\label{rapidity_sols}
\eta^{\pm}_{\nu} - \eta_l
= \pm \cosh^{-1}\!\left( 1 + \frac{\Delta^2}{m_W^2 - m_\perp^2} \right) 
\end{equation}
where 
\begin{equation}
\Delta^2 = \frac{m_W^2 - m_\perp^2}{2 p_\perp^\ell p_\perp^\nu}.
\end{equation}
and $m_\perp$ is the transverse mass of the $W$ boson.

The two-fold ambiguity in the neutrino rapidity propagates into an ambiguity in the azimuthal decay angle of the $W$ boson, $\phi$. This produces two solutions which, in the highly boosted limit, are related to each other by
\begin{equation}
\label{angle relation}
\phi^+ = (\pi - \phi^-) \mod 2\pi
\end{equation}
where $\phi^+$ and $\phi^-$ are the azimuthal decay angles obtained by taking the positive and negative rapidity solutions from Equation \ref{rapidity_sols} respectively.

\begin{figure}[H]
  \centering
  \includegraphics[width=0.48\textwidth]{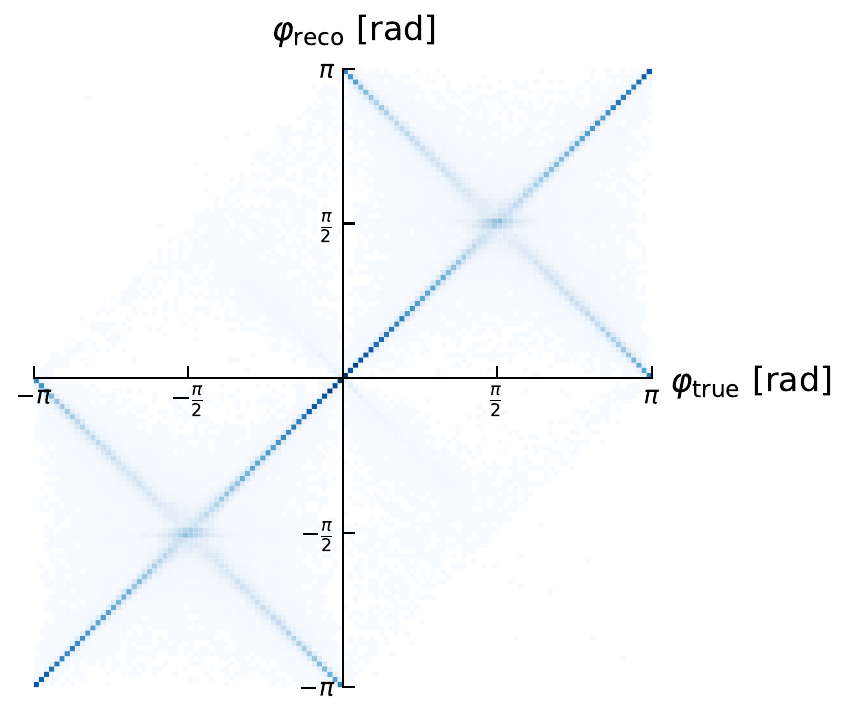}
  \caption{A heatmap of the reconstructed azimuthal decay angle $\phi_{reco}$ versus 
  the truth-level $\phi_{true}$ for the $W$ boson, with the neutrino rapidity solution 
  chosen at random from Eq.~(\ref{rapidity_sols}). The two diagonal bands reflect the 
  two-fold ambiguity of Eq.~(\ref{angle relation}), with each solution equally likely 
  to be selected.}
  \label{fig:neutrino_reco}
\end{figure}

\begin{figure}[H]
  \centering
  \includegraphics[width=0.48\textwidth]{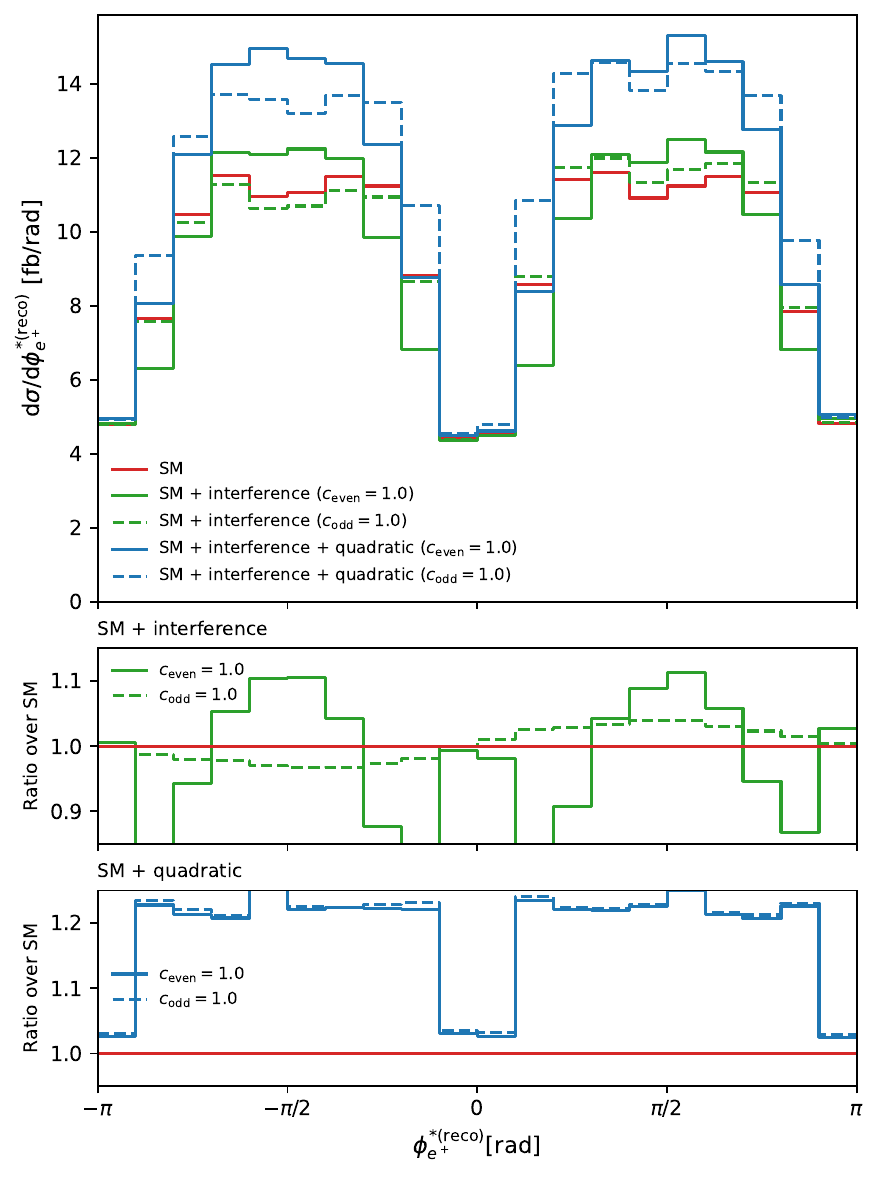}
  \caption{The $\phi_{reco}$ distribution at LO in QCD for the process 
  $pp \to W^+Z \to e^+ \nu_e \mu^+ \mu^-$ with the fiducial setup of Ref.~\cite{ElFaham:2024uop}, 
  applying the selections of Eq.~(3.2), with the neutrino rapidity solution 
  chosen at random from Eq.~(\ref{rapidity_sols}). 
  The upper panel shows the differential cross section $d\sigma/d\phi_e^*$ in units of fb, 
  comparing the SM prediction (red) with the SM + interference and SM + interference + quadratic 
  contributions for $c_{even} = 1.0$ (solid) and $c_{odd} = 1.0$ (dashed). 
  The middle and lower panels show the ratio to the SM for the interference-only 
  and quadratic contributions, respectively.}
  \label{fig:phi_neutrino_reco}
\end{figure}

Figure \ref{fig:neutrino_reco} shows the relationship between $\phi_{true}$, constructed using the truth-level neutrino rapidity, and $ \phi_{reco}\in \{\phi^+, \phi^-\}$, a solution of Equation \ref{rapidity_sols}, chosen at random. Figure~\ref{fig:phi_neutrino_reco} shows that the observable $\phi_{reco}$ retains some sensitivity to the EFT operators after the random choice of neutrino rapidity solution, though this is reduced. The overall effect of the fiducial cuts and the reconstruction ambiguity can be assessed by comparison with the inclusive truth-level result of Figure~\ref{fig:phi_neutrino_truth}.

As the functions $\sin2\phi$ and $\cos\phi$ change under the transformation \ref{angle relation}, 
the Fano coefficients, $a_i$, associated with Wigner P symbols 1, 5 and 6 can not be constrained due to the neutrino reconstruction ambiguity. Figures \ref{fig:cp_even_reco} and \ref{fig:cp_odd_reco} show the propagation of the neutrino reconstruction ambiguity into the spin density matrix for the quadratic term. 

\section{Sensitivity to New Physics}
To quantify sensitivity to new physics, we derive two-dimensional confidence intervals in the parameter space of Wilson coefficients~\cite{Wilson1969, Buchmuller1986}. We study the CP-even and CP-odd operators throughout this work~\cite{Bernreuther, Grzadkowski2010}, \(\left(c_W,\; c_{\widetilde{W}}\right),\) and compare the sensitivity obtained from three classes of observables: the azimuthal decay angle $\phi$ for the both the positron from the \(W\) and anti-muon from the \(Z\), $\phi_{e^+}^{*}$ and $\phi_{\mu^+}^{*}$~\cite{Dell_Aquila_1985, Bernreuther}, the transverse momentum of the decaying \(Z\) boson, $p_T^Z$, and the reconstructed spin density matrix~\cite{Ashby_Pickering_2023}. For each observable class we consider both a shape-only and a shape-plus-yield configuration, as described below. We expect for any chosen observable set, indexed by $\alpha$, the prediction may be expanded as
\begin{equation}
\begin{split}
\mu_\alpha(\vec{c}) =
\mu_\alpha^{\mathrm{SM}}&
+ c_W\,\mu_{\alpha,W}^{\mathrm{int}}
+ c_{\widetilde{W}}\,\mu_{\alpha,\widetilde{W}}^{\mathrm{int}}\\
&+ c_W^2\,\mu_{\alpha,W}^{\mathrm{quad}} 
+ c_{\widetilde{W}}^2\,\mu_{\alpha,\widetilde{W}}^{\mathrm{quad}}\\
&+ c_W c_{\widetilde{W}}\,\mu_{\alpha,W\widetilde{W}}^{\mathrm{mix}},
\end{split}
\label{eq:mu_expansion}
\end{equation}
where $\mu_\alpha^{\mathrm{SM}}$ denotes the SM expectation, the terms proportional to $c_i$ arise from SM--SMEFT interference~\cite{Brivio2019}, and the quadratic and mixed terms encode the self-interference and mutual interference of the two EFT operators.

Assuming Gaussian measurement uncertainties, we construct a $\chi^2$ test statistic 
of the form~\cite{Cowan2011}
\begin{equation}
\chi^2(\vec{c}) =
\sum_{\alpha,\beta}
\left[
x_\alpha - \mu_\alpha(\vec{c})
\right]
V^{-1}_{\alpha\beta}
\left[
x_\beta - \mu_\beta(\vec{c})
\right],
\label{eq:chi2_general}
\end{equation}
where $x_\alpha$ denotes the measured value of observable $\alpha$ and $V_{\alpha\beta}$ is the covariance matrix. Expected limits are obtained under the Asimov assumption~\cite{Cowan2011} at luminosity $\mathcal{L} = 140~\text{fb}^{-1}$, with the data vector set equal to the SM prediction,
\[
x_\alpha = \mu_\alpha^{\mathrm{SM}},
\]
so that Eq.~\eqref{eq:chi2_general} reduces to a measure of the distance between the EFT hypothesis and the SM expectation. The allowed region at confidence level $1-\alpha$ 
is defined by
\begin{equation}
\Delta \chi^2(\vec{c}) \equiv
\chi^2_{\mathrm{A}}(\vec{c}) - \chi^2_{\mathrm{A}}(\hat{\vec{c}})
\leq \Delta\chi^2_{2,\alpha},
\label{eq:deltachi2_def}
\end{equation}
where $\hat{\vec{c}} = (0,0)$ at the Asimov SM point, so that $\chi^2_{\mathrm{A}}(\hat{\vec{c}}) = 0$. For two parameters of interest, the standard asymptotic thresholds are~\cite{Wilks1938, Cowan2011}
\begin{equation}
\Delta\chi^2_{2,68\%} = 2.30,
\qquad
\Delta\chi^2_{2,95\%} = 5.99.
\end{equation}
Contours satisfying these conditions define the joint confidence intervals shown in Figures~\ref{fig:contours}, ~\ref{fig:contours_fiducial_phis} and ~\ref{fig:contours_fiducial_ptz_sdm}.

\subsection{Confidence Limits from the \texorpdfstring{$\phi$}{phi} and \texorpdfstring{$p_T^Z$}{pT of Z} Distributions}

The azimuthal decay angle $\phi$ has previously been shown to resurrect the interference effects that are suppressed in polarisation-inclusive observables at high energy~\cite{Panico_2018, Banerjee:2019twi}. The index $\alpha$ in Eq.~\eqref{eq:mu_expansion} labels bins of the $\phi$ distribution.

For the shape-only configuration, the covariance is the multinomial variance of the normalised bin fractions evaluated at the SM hypothesis,
\begin{equation}
V_{\alpha\beta}^{(\phi)} = \frac{1}{N_{\rm SM}}
\left(\delta_{\alpha\beta}\, p_\alpha - p_\alpha p_\beta\right),
\label{eq:phi_cov}
\end{equation}
where $p_\alpha$ are the SM bin fractions and $N_{\rm SM} = \mathcal{L}\,\sigma_{\rm SM}$. The last bin is dropped to respect the constraint $\sum_\alpha p_\alpha = 1$. For the shape-plus-yield configuration, the full yield vector $y_\alpha = \mathcal{L}\,\sigma_\alpha(\vec{c})$ is used with a diagonal Poisson covariance $V_{\alpha\beta} = \delta_{\alpha\beta}\,y_\alpha^{\rm SM}$, which is equivalent to adding an independent rate $\chi^2$ contribution to the shape statistic.

The shape-only statistic is conditional on the expected event count, since $N_{\rm SM}$ sets the statistical precision, but it does not use deviations in the total yield as an observable. The two configurations should therefore be interpreted as reflecting the sensitivity from angular modulations alone versus angular modulations combined with the total event yield, rather than as strictly orthogonal contributions. Extracting limits using both the $\phi_{e^+}^{*}$ and $\phi_{\mu^+}^{*}$ angles allows for insight into the diminishing effect caused by neutrino reconstruction has on the former, which the latter does not suffer from.

Following this procedure we also include $p_T^Z$ as a separate observable for limit extraction, allowing a comparison to a purely kinematic variable with differential tail sensitivity to the SMEFT operators cross-section enhancement, while having limited interference resurrection power.

\subsection{Confidence Limits from the Spin Density Matrix}

The central result of this work is obtained by constructing the test statistic directly from the reconstructed spin density matrix. The observable vector is built from the independent real and imaginary components of the Wigner-P function means,
\begin{equation}
x_\alpha \in \{a_i,\; b_j,\; c_{ij}\},
\end{equation}
as defined in Eqs.~\eqref{eq:WPAverage} and~\eqref{eq:c_ij}. Since the SDM contains both single-boson polarisation information and genuine spin-correlation information, it retains more of the structure of the underlying helicity amplitudes than one-dimensional 
angular observables alone.

The event-level covariance $\Sigma_{\rm event}$ is estimated from the second moment of the Wigner-P feature distribution,
\begin{equation}
\Sigma_{\rm event} = \left\langle \vec{f}\,\vec{f}^{\,T} \right\rangle - 
\vec{\mu}\,\vec{\mu}^{\,T},
\end{equation}
where $\vec{f}$ denotes the feature vector for a single event. The covariance of the 
sample mean at the SM point is then $V = \Sigma_{\rm event}^{\rm SM} / N_{\rm SM}$, 
and the test statistic is
\begin{equation}
\chi^2_{\mathrm{SDM}}(\vec{c}) =
\left[\vec{\mu}(\vec{c}) - \vec{\mu}^{\,\rm SM}\right]^T
V^{-1}
\left[\vec{\mu}(\vec{c}) - \vec{\mu}^{\,\rm SM}\right].
\label{eq:chi2_sdm}
\end{equation}
As with the $\phi$ case, this is a shape-only statistic up to the implicit rate dependence through $N_{\rm SM}$ in the covariance, and the shape-plus-yield variant adds an independent yield $\chi^2$ term. A regularisation $\epsilon\mathbf{I}$ with $\epsilon = 10^{-8}$ is added to all covariance matrices prior to inversion to ensure numerical stability.

The sensitivity of this construction can already be anticipated from Figures~\ref{fig:int_term} and~\ref{fig:quadratic_term}. In the interference term, the CP-even and CP-odd operators populate qualitatively different parts of the SDM: $O_W$ contributes predominantly to the real components, while $O_{\widetilde{W}}$ generates off-diagonal imaginary structure, consistent with Eq.~\eqref{eq:SDMAmp}. In the quadratic term, the two operators produce distinct off-diagonal patterns even though they are known to be degenerate in traditional angular observables~\cite{ElFaham:2024uop}. The SDM therefore provides a higher-dimensional observable space in which these operator degeneracies are reduced, going beyond mere interference resurrection~\cite{Panico_2018} to exploit the full spin structure of the diboson final state.

We also extract one-dimensional confidence intervals on each Wilson coefficient individually. Two procedures are considered. In the \emph{profiled} approach, the test statistic is minimised over the other operator at each scan point,
\begin{equation}
\Delta\chi^2_{\rm prof}(c_i) = \min_{c_j}\,\chi^2_{\mathrm{A}}(c_i, c_j),
\end{equation}
yielding intervals that marginalise over the unknown value of the second coefficient. In the \emph{fixed} approach, the other operator is held at its SM value $c_j = 0$, giving the sensitivity that would be obtained if only one operator were present. Since the test statistic now depends on a single parameter, the asymptotic thresholds for the $68\%$ and $95\%$ confidence levels are $\Delta\chi^2_{1,68\%} = 1.00$ and $\Delta\chi^2_{1,95\%} = 3.84$, respectively~\cite{Wilks1938, Cowan2011}. 
\begin{figure}[H]
  \centering
  \includegraphics[width=0.48\textwidth]{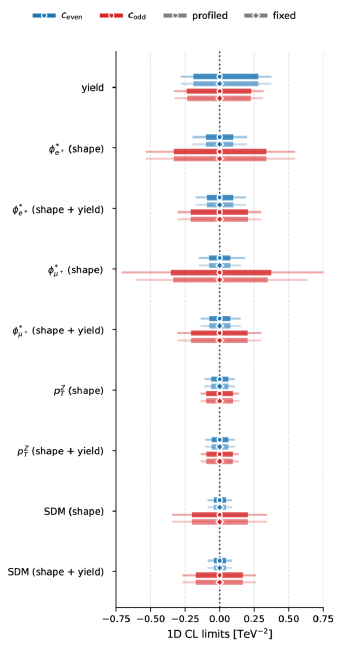}
  \caption{1D CL limits on $O_W$ and $O_{\widetilde{W}}$ derived with other operator fixed to zero  
  and with the other operator profiled over, for the shape-only and shape-plus-yield 
  configurations of the $\phi$, $p_T^Z$ and SDM observables. The boxes represent 68\% CL limit bounds, the lines represent 95\% CL limit bounds. The limits are produced within the fiducial selection.}
  \label{fig:1d_limits}
\end{figure}
The resulting confidence limits are shown in Figure~\ref{fig:1d_limits} for each observable configuration. The difference between the profiled and fixed intervals reflects the degree of correlation between $c_W$ and $c_{\widetilde{W}}$ in a given observable: where the two procedures yield similar intervals, the observable effectively decouples the two operators. As the cross-term between the $O_W$ and $O_{\widetilde{W}}$ operators is negligibly small and the operators populate largely orthogonal parts of the observable space, the profiled limits are only marginally weaker than the fixed approach, as expected. Of the simple kinematic observables, $p_T^Z$ provides the strongest limits on both operators. However the SDM retains competitiveness when combined with cross-section information, providing the next strongest 1D bounds on the Wilson coefficient of $O_{\widetilde{W}}$ operator after the $p_T^Z$ constraints and producing the strongest limits on the Wilson coefficient of the CP-even $O_W$ operator.

\subsection[Two-dimensional Limits on W Operators]{Two-Dimensional Limits on $O_W$ and $O_{\widetilde{W}}$}

\begin{figure*}[p]
    \centering
    \begin{subfigure}{0.49\textwidth}
        \centering
        \includegraphics[width=\linewidth]{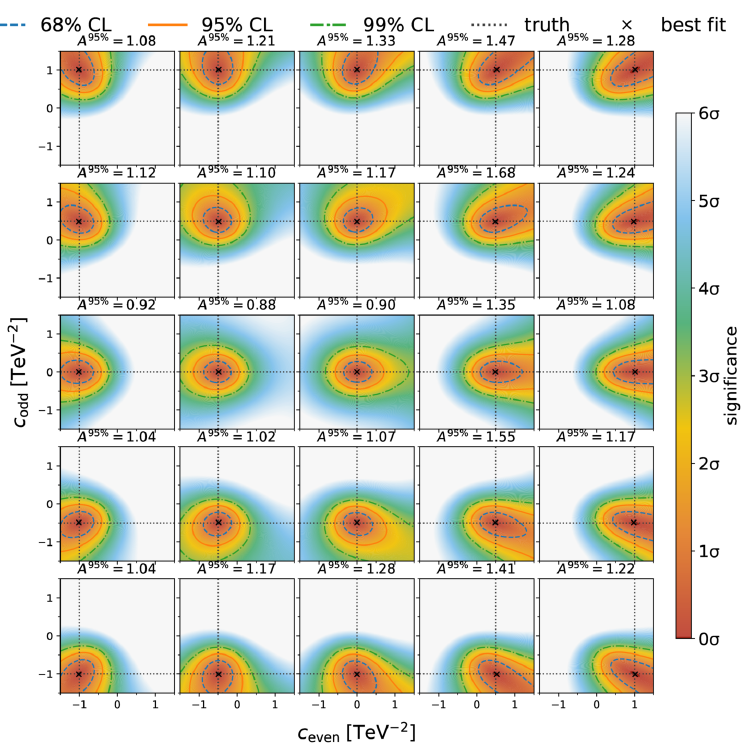}
        \caption{Inclusive $\phi_{e^+}^{*}$ constrained limits (shape-only)}
        \label{fig:phi_contours}
    \end{subfigure}
    \hfill
    \begin{subfigure}{0.49\textwidth}
        \centering
        \includegraphics[width=\linewidth]{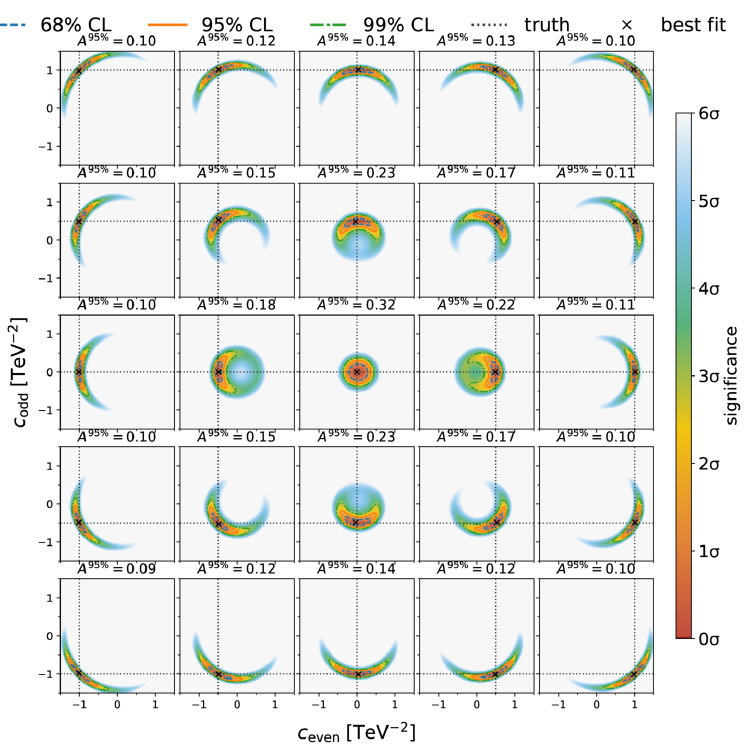}
        \caption{Inclusive $\phi_{e^+}^{*}$ constrained limits (shape + yield)}
        \label{fig:phiandxsec_contours}
    \end{subfigure}
    \vspace{6em}
    \begin{subfigure}{0.49\textwidth}
        \centering
        \includegraphics[width=\linewidth]{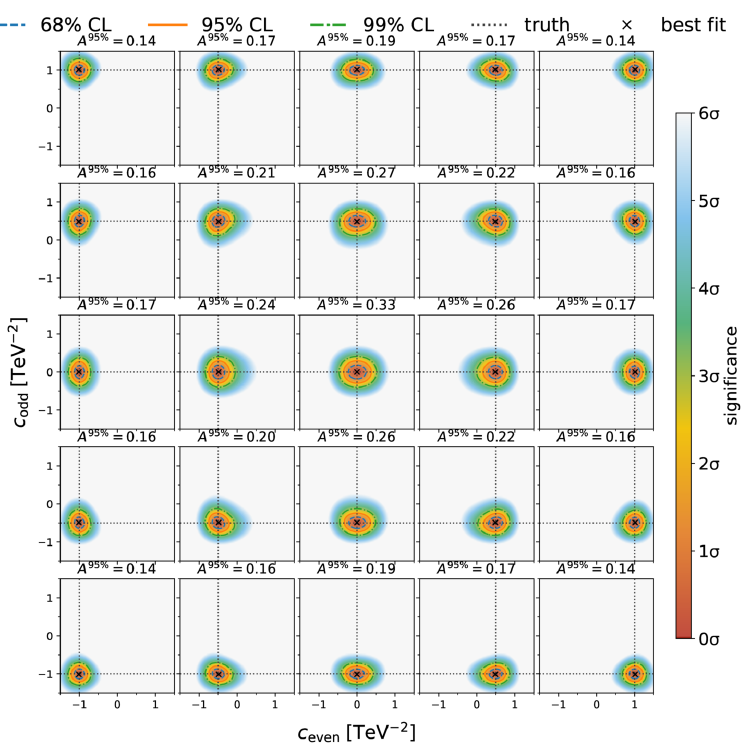}
        \caption{Inclusive SDM constrained limits (shape-only)}
        \label{fig:sdm_contours}
    \end{subfigure}
    \hfill
    \begin{subfigure}{0.49\textwidth}
        \centering
        \includegraphics[width=\linewidth]{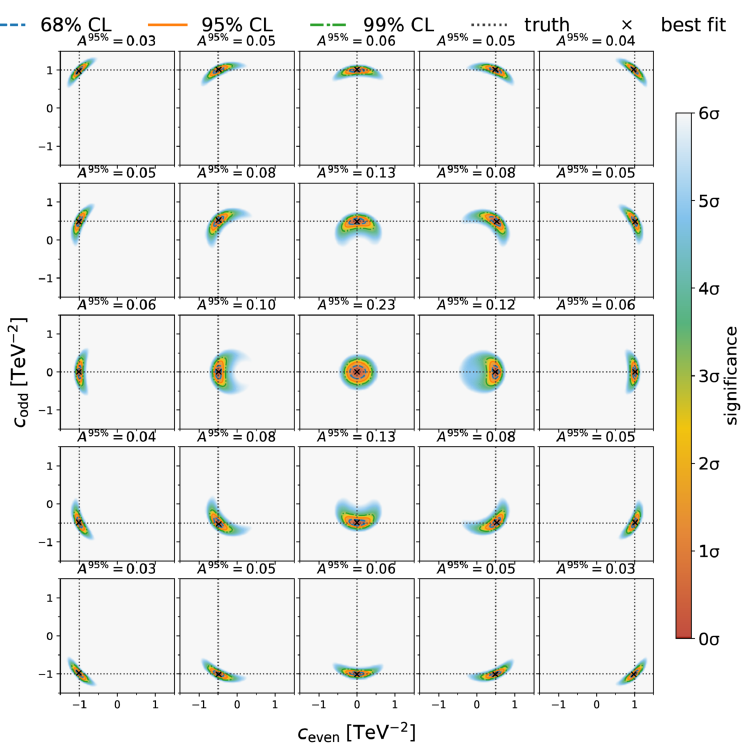}
        \caption{Inclusive SDM constrained limits (shape + yield)}
        \label{fig:sdmandxsec_contours}
    \end{subfigure}
    \caption{Expected joint confidence regions in the $(c_W,c_{\widetilde{W}})$ plane derived from the positron azimuthal decay-angle distribution $\phi_{e^+}^{*}$ (top) and the reconstructed spin density matrix (bottom), each in shape-only (left) and shape-plus-yield (right) configurations, for the inclusive setup. The contours correspond to constant $\Delta\chi^2$ with respect to the SM point at fixed confidence levels. Contour limits are extracted based on both a SM Asimov scenario, but also a set of NP Asimov setups with varying degrees of $O_W$ and $O_{\widetilde{W}}$ present in the dataset across the grid. This demonstrates of the SDM-based analysis's ability to differentiate actual CP-even and CP-odd signatures from each other unambiguously. The SDM analysis provides the stronger separation between $O_W$ and $O_{\widetilde{W}}$, reflecting the additional spin-correlation information retained by the tomographic reconstruction. This is reflected in the total areas of the 95\% CL contours $A^{\textrm{95\%}}$, which are shown above each subplot of the Asimov grids.}
    \label{fig:contours}
\end{figure*}

\begin{figure*}[p]
    \centering
    \begin{subfigure}{0.49\textwidth}
        \centering
        \includegraphics[width=\linewidth]{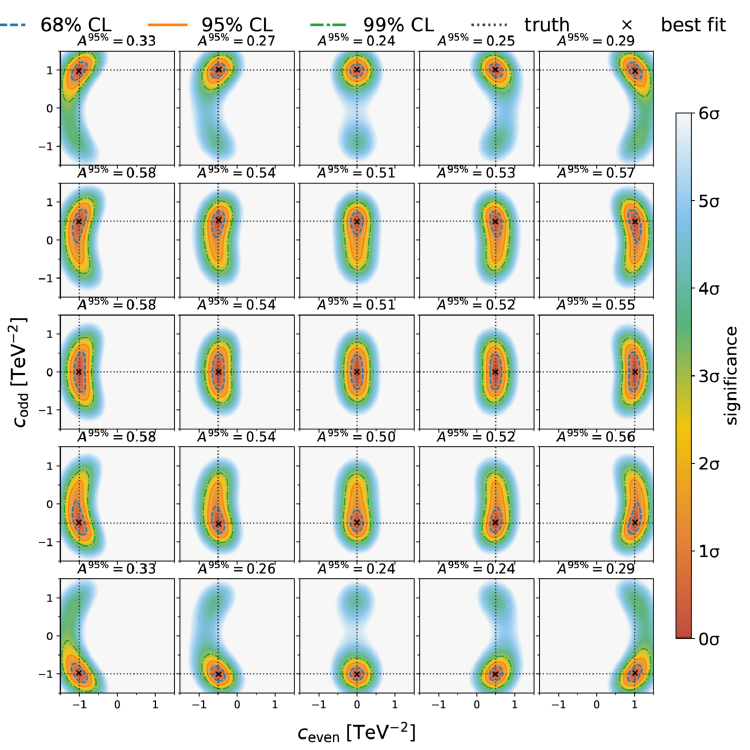}
        \caption{Fiducial $\phi_{e^+}^{*}$ constrained limits (shape-only)}
        \label{fig:phi_contours_fiducial}
    \end{subfigure}
    \hfill
    \begin{subfigure}{0.49\textwidth}
        \centering
        \includegraphics[width=\linewidth]{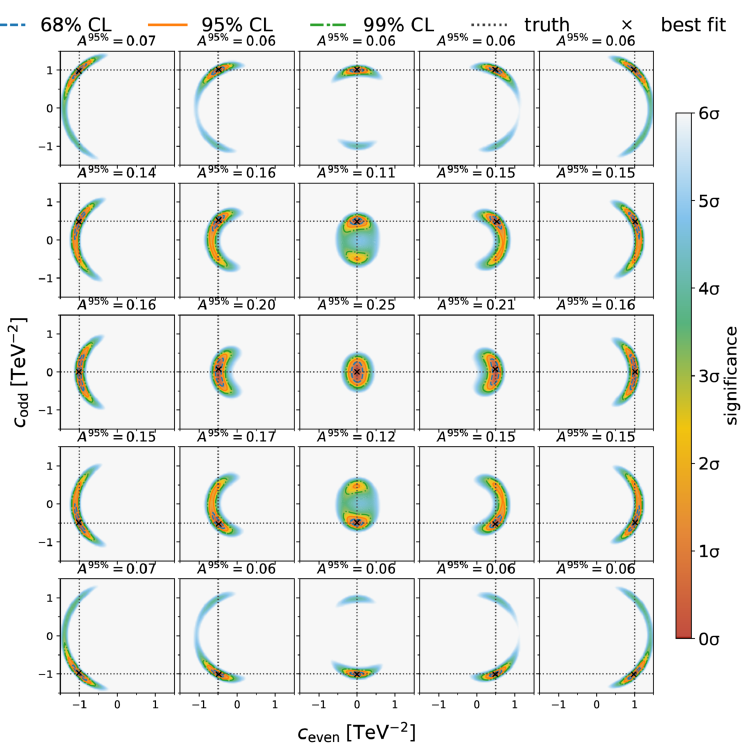}
        \caption{Fiducial $\phi_{e^+}^{*}$ constrained limits (shape + yield)}
        \label{fig:phiandxsec_contours_fiducial}
    \end{subfigure}
    \vspace{6em}
    \begin{subfigure}{0.49\textwidth}
        \centering
        \includegraphics[width=\linewidth]{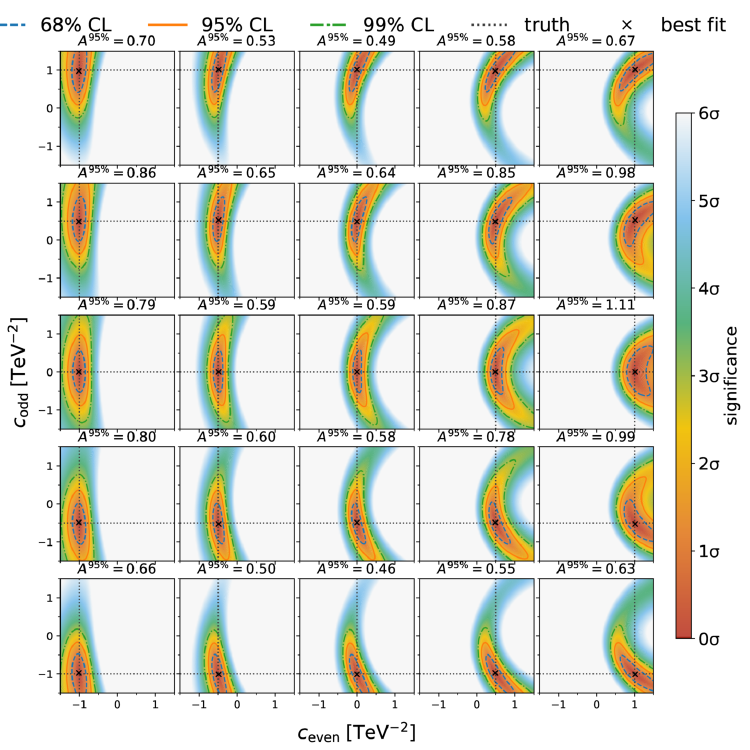}
        \caption{Fiducial $\phi_{\mu^+}^{*}$ constrained limits (shape-only)}
        \label{fig:muphi_contours_fiducial}
    \end{subfigure}
    \hfill
    \begin{subfigure}{0.49\textwidth}
        \centering
        \includegraphics[width=\linewidth]{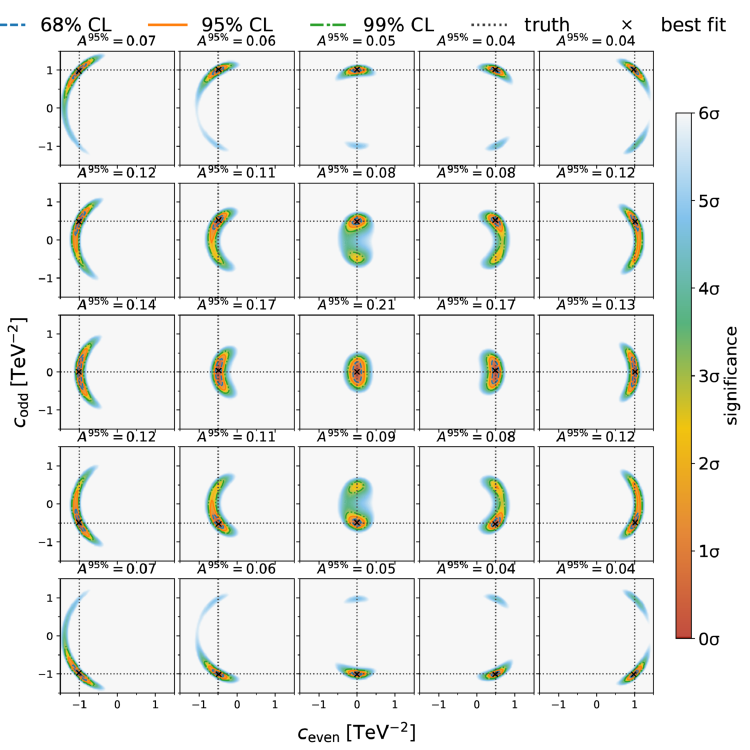}
        \caption{Fiducial $\phi_{\mu^+}^{*}$ constrained limits (shape + yield)}
        \label{fig:muphiandxsec_contours_fiducial}
    \end{subfigure}
    \caption{Expected joint confidence regions in the $(c_W,c_{\widetilde{W}})$ plane derived from the positron and anti-muon azimuthal decay-angle distributions, $\phi_{e^+}^{*}$ (top) and $\phi_{\mu^+}^{*}$ with the fiducial selection of Eq.~\eqref{eq:fiducial} applied and the neutrino rapidity reconstructed via the on-shell $W$-mass constraint of Eq.~(\ref{rapidity_sols}). The effect of the fiducial cuts and reconstruction ambiguity on the sensitivity of $\phi_{e^+}^{*}$ can be assessed by comparison with Figures~\ref{fig:phi_contours} and ~\ref{fig:phiandxsec_contours}. Constraints from $\phi_{\mu^+}^{*}$ are unaffected by neutrino reconstruction and so slightly exceed the strength of those from $\phi_{e^+}^{*}$.}
    \label{fig:contours_fiducial_phis}
\end{figure*}

\begin{figure*}[p]
    \centering
    \begin{subfigure}{0.49\textwidth}
        \centering
        \includegraphics[width=\linewidth]{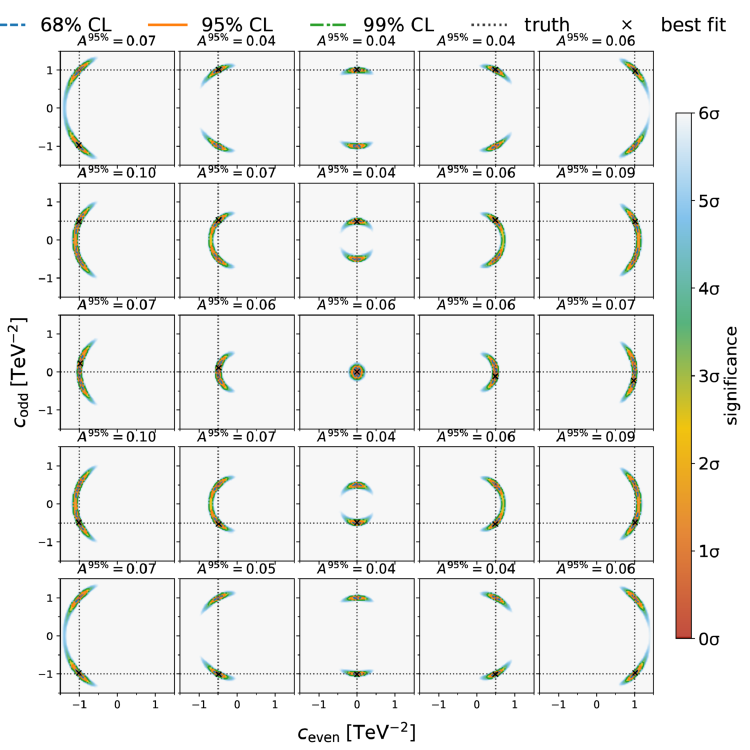}
        \caption{Fiducial $p_T^Z$ constrained limits (shape-only)}
        \label{fig:ptz_contours_fiducial}
    \end{subfigure}
    \hfill
    \begin{subfigure}{0.49\textwidth}
        \centering
        \includegraphics[width=\linewidth]{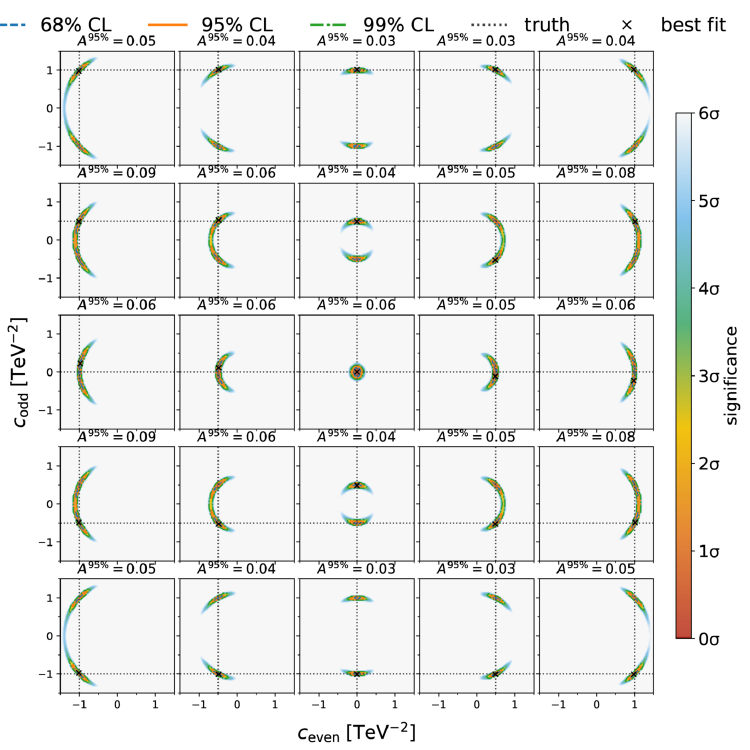}
        \caption{Fiducial $p_T^Z$ constrained limits (shape + yield)}
        \label{fig:ptzandxsec_contours_fiducial}
    \end{subfigure}
    \vspace{6em}
    \begin{subfigure}{0.49\textwidth}
        \centering
        \includegraphics[width=\linewidth]{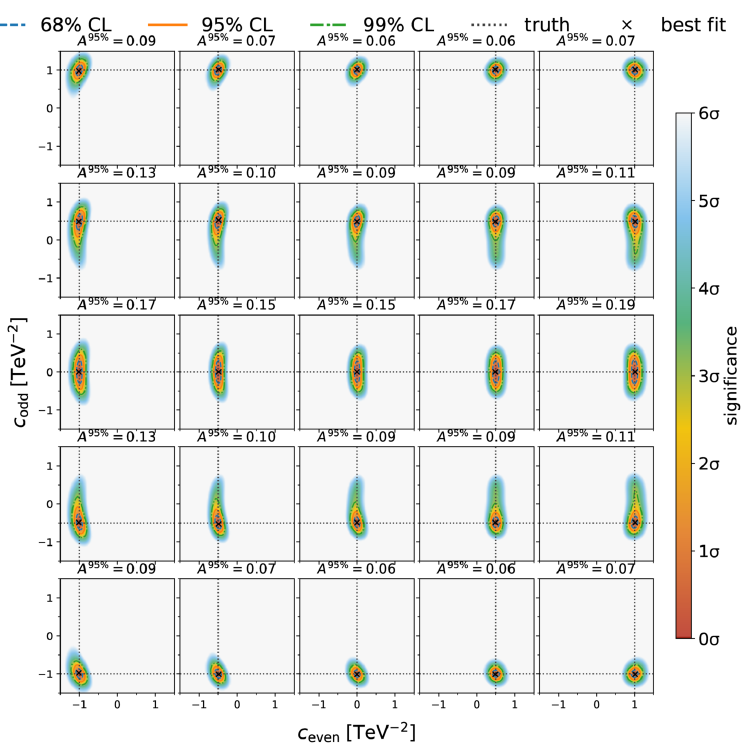}
        \caption{Fiducial SDM constrained limits (shape-only)}
        \label{fig:sdm_contours_fiducial}
    \end{subfigure}
    \hfill
    \begin{subfigure}{0.49\textwidth}
        \centering
        \includegraphics[width=\linewidth]{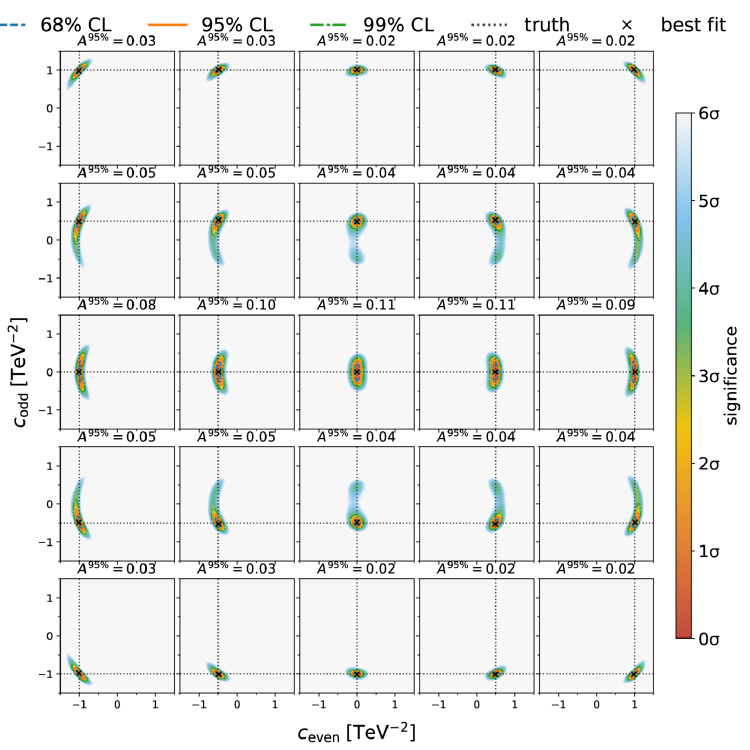}
        \caption{Fiducial SDM constrained limits (shape + yield)}
        \label{fig:sdmandxsec_contours_fiducial}
    \end{subfigure}
    \caption{Expected joint confidence regions in the $(c_W,c_{\widetilde{W}})$ plane derived from the positron and anti-muon azimuthal decay-angle distributions, $p_T^Z$ (top) and the SDM (bottom) with the fiducial selection of Eq.~\eqref{eq:fiducial} applied and the neutrino rapidity reconstructed. The differential nature of the $p_T^Z$ distribution tail gives strong constraints of the SM Asimov, with a smaller contour area, $A^{\textrm{95\%}}$ than achieved by the SDM with the addition of cross-section information. The SDM however performs better by the same metric in the genuine BSM scenarios presented in the off-centre points of the Asimov grid. In these cases the SDM also show clearer differentiation between BSM hypotheses than $p_T^Z$, with more accurate assignment of the best fit value to the truth value of the Asimov.}
    \label{fig:contours_fiducial_ptz_sdm}
\end{figure*}

The resulting two-dimensional confidence regions in the $(c_W,c_{\widetilde{W}})$ plane are shown in Figure~\ref{fig:contours} for the inclusive setup and Figures~\ref{fig:contours_fiducial_phis} and~\ref{fig:contours_fiducial_ptz_sdm} for the fiducial setup with neutrino reconstruction. In each case, Figures~\ref{fig:phi_contours} and~\ref{fig:phi_contours_fiducial} show the $\phi_{e^+}^{*}$-only limits, Figures~\ref{fig:phiandxsec_contours} and~\ref{fig:phiandxsec_contours_fiducial} the shape-plus-yield $\phi_{e^+}^{*}$ limits. Figure~\ref{fig:muphi_contours_fiducial} show the $\phi_{\mu^+}^{*}$-only limits, Figure~\ref{fig:muphiandxsec_contours_fiducial} the shape-plus-yield $\phi_{\mu^+}^{*}$ limits, both in the fiducial setup. Figure~\ref{fig:ptz_contours_fiducial} show the $p_T^Z$-only limits, Figure~\ref{fig:ptzandxsec_contours_fiducial} the shape-plus-yield $p_T^Z$ limits, both in the fiducial setup. Figures~\ref{fig:sdm_contours} and~\ref{fig:sdm_contours_fiducial} the SDM shape-only limits, and Figures~\ref{fig:sdmandxsec_contours} and~\ref{fig:sdmandxsec_contours_fiducial} the SDM shape-plus-yield limits.

The $\phi_{e^+}^{*}$ and $\phi_{\mu^+}^{*}$ observables show linear sensitivity to new physics through SM--EFT interference, improving the separation between CP-even and CP-odd contributions relative to a cross-section only measurement. However, because they probe only a restricted projection of the available spin information, residual degeneracies between $O_W$ and $O_{\widetilde{W}}$ remain, particularly in 
the quadratic regime. 

The observable $p_T^Z$ enables tighter two dimensional limits to be obtained than the $\phi_{e^+}^{*}$ and $\phi_{\mu^+}^{*}$ observables and is able to obtain a tighter two dimensional bound on the SM in the absence of BSM physics, i.e. in the central contour plot. However, using  $p_T^Z$ alone, the two-fold degeneracy between odd and even new physics signatures remains at a $5\sigma$ confidence level, as well insensitivity to the sign of each Wilson coefficient. 

The SDM retains not only angular modulation information but also the full pattern of spin correlations encoded in the $c_{ij}$ matrix of Eq.~\eqref{eq:DB_Rho}. In particular, the off-diagonal matrix elements provide direct sensitivity to structures that distinguish CP-even from CP-odd dynamics, yielding tighter and less degenerate joint confidence intervals on $(c_W, c_{\widetilde{W}})$.

\section{Conclusions}

We have studied the sensitivity of quantum tomography to CP-even and CP-odd dimension-six SMEFT effects in the process
\[
pp \to W^+Z \to e^+ \nu_e \mu^+ \mu^-,
\]
with particular focus on the operators $O_W$ and $O_{\widetilde{W}}$. The motivation for this study is that conventional polarisation-inclusive observables are known to lose linear sensitivity to these effects at high energy, while standard angular observables such as the azimuthal decay angle $\phi$ recover only part of the information contained in the diboson final state.

By reconstructing the spin density matrix of the $W^+Z$ system, we have shown that quantum tomography provides access to the full spin structure of the process. This allows us to distinguish CP-even and CP-odd SMEFT contributions in a more complete way than traditional one-dimensional observables. In particular, the interference terms of the two operators populate different components of the SDM: the CP-even contribution appears predominantly in the real part, while the CP-odd contribution generates characteristic imaginary off-diagonal structure. Beyond interference resurrection, the SDM also retains discriminating power in the quadratic SMEFT terms, where the usual $\phi$ distributions become largely degenerate and kinematic observables such as $p_T^Z$ show little differentiating power between the operators.

We also investigated the impact of neutrino reconstruction in a fiducial setup relevant for experiment. As expected, the two-fold ambiguity in the neutrino degrades sensitivity to a subset of SDM components and weakens the discrimination power of observables constructed from the $W$ decay angles. Nevertheless, the reconstructed SDM remains useful information and continues to separate CP-even and CP-odd effects effectively. This demonstrates that the tomographic approach is robust enough to remain phenomenologically useful in more realistic leptonic $WZ$ analyses.

Using SM and BSM Asimov likelihoods in the $(c_W,c_{\widetilde{W}})$ plane, we compared the expected sensitivity of the reconstructed SDM against the azimuthal decay angles $\phi_{e^+}^{*}$ and $\phi_{\mu^+}^{*}$ and the kinematic observable $p_T^Z$, considering both shape-only and shape-plus-yield configurations. We found that the SDM provides a clear simultaneous separation of CP-even and CP-odd effects, producing less degenerate confidence regions than the angular observables. 

Kinematic variables remain important for constraining the operators we examine in this work, as we find $p_T^Z$ to produce more stringent one-dimensional limits on the CP-odd $O_{\widetilde{W}}$ operator, and tighter two dimensional contour bounds in the SM Asimov data compared to the SDM. However, the SDM approach presented here exceeds the one-dimensional constraining power of $p_T^Z$ for the CP-even $O_W$ operator, and provides superior differentiating power scenarios where genuine BSM physics is present, making the SDM a powerful tool for potential discoveries at the LHC and beyond. The methodology shown here can be expanded  further, to utilise kinematic information in order to use the SDM differentially or to isolate a NP-enhanced region of phase space. Limits from kinematic observables are also more sensitive to unitarity bounds, which are not present in this study, such as those applied in ATLAS's interpretation of their recent $W^+Z$ production measurement~\cite{ATLAS:2025edf}. The application of clipping scans in future studies to remove unphysical unitarity-violating contributions from the SMEFT operators could greatly benefit the competitiveness of the SDM approach.

The present study is intentionally idealised. Background contamination, detector effects, and systematic uncertainties were neglected in order to isolate the intrinsic sensitivity of the SDM to SMEFT effects. In addition, all results were obtained at leading order, while higher-order QCD and electroweak corrections are known to be important in $WZ$ production. Extending the tomographic framework to include these effects, together with a full treatment of experimental systematics and additional diboson channels, is therefore an important next step.

Overall, our results show that quantum tomography offers a powerful framework for SMEFT studies in diboson production. Rather than relying only on interference resurrection in selected angular observables, or cross-section enhancement in the tails of kinematic observables, the SDM approach exploits the full observable imprint of the helicity amplitudes, including information that survives in the quadratic SMEFT regime. This makes it a promising tool for future LHC analyses aiming to constrain or disentangle new CP-violating interactions in the electroweak sector, and for potentially discovering physics beyond the SM.

\paragraph{Acknowledgments:} We would like to thank Andy Pilkington, Eleni Vryonidou and Christoph Englert for the helpful discussions had during the early development of this work. P.D. thanks Saptaparna Bhattacharya for her support of this project. A.R. is supported by a PhD studentship at the University of Manchester through the UK Science and Technology Facilities Council (STFC) under grant ST/Y509814/1. P.D. is supported by Southern Methodist University through Prof. S. Bhattacharya's faculty startup funding. A.O. and S.S. are supported under the Science and Technology Facilities Council grant ST/W000601/1.

\appendix

\section{Numerical One-Dimensional Limits}

\begin{table}[H]
\centering
\caption{1D profiled 95\% CL intervals on $c_W$ and $c_{\widetilde{W}}$ at
$(c_W, c_{\widetilde{W}}) = (0, 0)$. The other operator is profiled over
in each case. Limits are extracted in the fiducial setup.}
\label{tab:limits_profiled}
\begin{adjustbox}{width=\columnwidth}
\begin{tabular}{l cc}
\hline\hline
Mode & $c_W$ 95\% CL & $c_{\widetilde{W}}$ 95\% CL \\
\hline
Yield
  & $[-0.282,\;+0.373]$ & $[-0.330,\;+0.317]$ \\
$\phi_{e^+}^{*}$ (shape)
  & $[-0.197,\;+0.197]$ & $[-0.535,\;+0.544]$ \\
$\phi_{e^+}^{*}$ (shape + yield)
  & $[-0.175,\;+0.192]$ & $[-0.307,\;+0.298]$ \\
$\phi_{\mu^+}^{*}$ (shape)
  & $[-0.150,\;+0.186]$ & $[-0.712,\;+0.784]$ \\
$\phi_{\mu^+}^{*}$ (shape + yield)
  & $[-0.140,\;+0.152]$ & $[-0.310,\;+0.302]$ \\
$p_T^Z$ (shape)
  & $[-0.109,\;+0.112]$ & $[-0.141,\;+0.072]$ \\
$p_T^Z$ (shape + yield)
  & $[-0.106,\;+0.112]$ & $[-0.139,\;+0.138]$ \\
SDM (shape)
  & $[-0.088,\;+0.089]$ & $[-0.342,\;+0.342]$ \\
SDM (shape + yield)
  & $[-0.087,\;+0.089]$ & $[-0.270,\;+0.264]$ \\
\hline\hline
\end{tabular}
\end{adjustbox}
\end{table}    

\begin{table}[H]
\centering
\caption{1D fixed 95\% CL intervals on $c_W$ and $c_{\widetilde{W}}$ at
$(c_W, c_{\widetilde{W}}) = (0, 0)$. The other operator is fixed to zero. Limits are extracted in the fiducial setup.}
\label{tab:limits_fixed}
\begin{adjustbox}{width=\columnwidth}
\begin{tabular}{l cc}
\hline\hline
Mode & $c_W$ 95\% CL & $c_{\widetilde{W}}$ 95\% CL \\
\hline
Yield
  & $[-0.282,\;+0.373]$ & $[-0.326,\;+0.314]$ \\
$\phi_{e^+}^{*}$ (shape)
  & $[-0.197,\;+0.197]$ & $[-0.535,\;+0.544]$ \\
$\phi_{e^+}^{*}$ (shape + yield)
  & $[-0.175,\;+0.192]$ & $[-0.307,\;+0.298]$ \\
$\phi_{\mu^+}^{*}$ (shape)
  & $[-0.150,\;+0.153]$ & $[-0.603,\;+0.633]$ \\
$\phi_{\mu^+}^{*}$ (shape + yield)
  & $[-0.140,\;+0.152]$ & $[-0.307,\;+0.299]$ \\
$p_T^Z$ (shape)
  & $[-0.109,\;+0.112]$ & $[-0.141,\;+0.140]$ \\
$p_T^Z$ (shape + yield)
  & $[-0.106,\;+0.112]$ & $[-0.139,\;+0.138]$ \\
SDM (shape)
  & $[-0.088,\;+0.089]$ & $[-0.341,\;+0.342]$ \\
SDM (shape + yield)
  & $[-0.087,\;+0.089]$ & $[-0.270,\;+0.264]$ \\
\hline\hline
\end{tabular}
\end{adjustbox}
\end{table} 

\section{Transformation Matrix for the Z Boson}
\label{sec:a_matrix}
The transformation matrix $A$ used in Eq.~\eqref{eq:generalized_p_z} to construct 
the generalised Wigner P-symbols for the $Z \to \ell^+\ell^-$ decay is given by

\begin{equation}
\begingroup
\setlength{\arraycolsep}{0.25pt}
\resizebox{\columnwidth}{!}{$
A = \begin{pmatrix}
|c_R|^2 & 0 & 0 & 0 & 0 & |c_L|^2 & 0 & 0 \\
0 & |c_R|^2 & 0 & 0 & 0 & 0 & |c_L|^2 & 0 \\
0 & 0 & \tfrac{1}{2}|c_R|^2 - \tfrac{1}{2}|c_L|^2 & 0 & 0 & 0 & 0 & \tfrac{\sqrt{3}}{2}|c_L|^2 \\
0 & 0 & 0 & |c_R|^2 - |c_L|^2 & 0 & 0 & 0 & 0 \\
0 & 0 & 0 & 0 & |c_R|^2 - |c_L|^2 & 0 & 0 & 0 \\
|c_L|^2 & 0 & 0 & 0 & 0 & |c_R|^2 & 0 & 0 \\
0 & |c_L|^2 & 0 & 0 & 0 & 0 & |c_R|^2 & 0 \\
0 & 0 & \tfrac{\sqrt{3}}{2}|c_L|^2 & 0 & 0 & 0 & 0 &
\tfrac{1}{2}|c_L|^2 + |c_R|^2
\end{pmatrix}
$}
\endgroup
\label{eq:A_matrix}
\end{equation}
where $c_L = -0.273$ and $c_R = +0.233$ are the left- and right-handed couplings 
of the Z boson to leptons~\cite{Ashby_Pickering_2023}.

\section{Reconstructed Gell Mann Parameters of the WZ process}

\begin{figure}[H]
  \centering
  \includegraphics[width=0.8\linewidth]{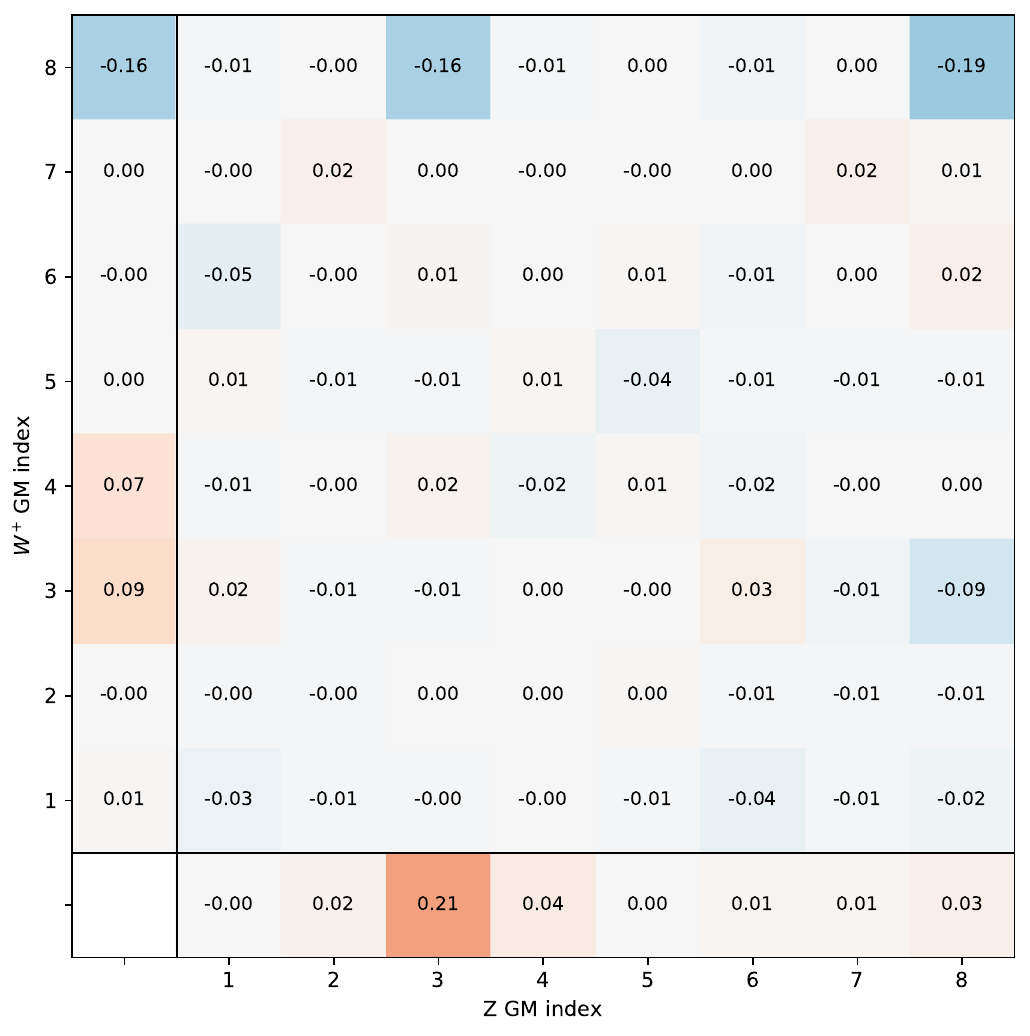}
  \caption{Gell-Mann coefficient matrix for the \(W^+Z\) system in the SM. 
  The left column and bottom row correspond to the single-boson polarization vectors \(a_i\) and \(b_j\), respectively, while the central \(8 \times 8\) block shows the spin correlation coefficients \(c_{ij}\). The bottom-left square has no physical meaning.}
  \label{fig:wz_cij_sm}
\end{figure}

\begin{figure}[H]
  \centering
  \includegraphics[width=0.8\linewidth]{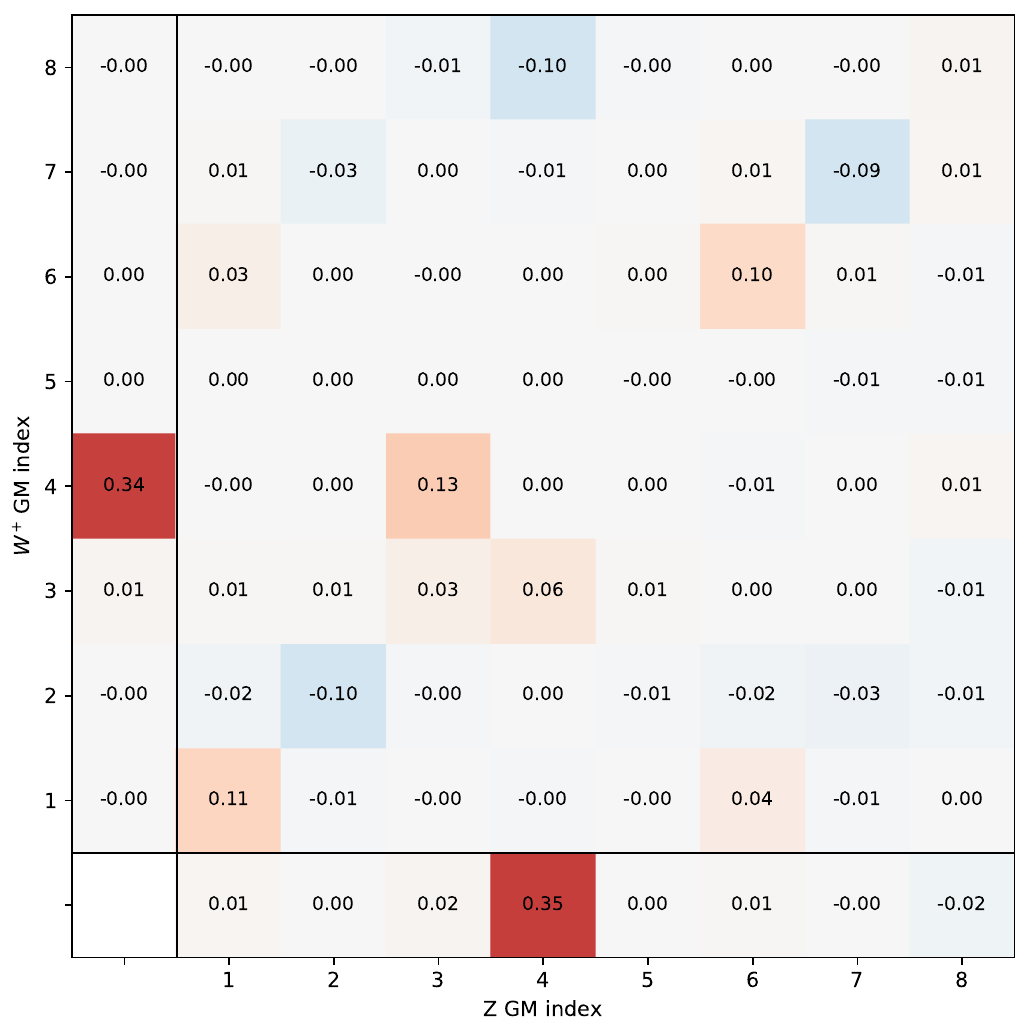}
  \caption{Gell-Mann coefficient matrix for the \(W^+Z\) system: even interference contribution.}
  \label{fig:wz_cij_even_int}
\end{figure}

\begin{figure}[H]
  \centering
  \includegraphics[width=0.8\linewidth]{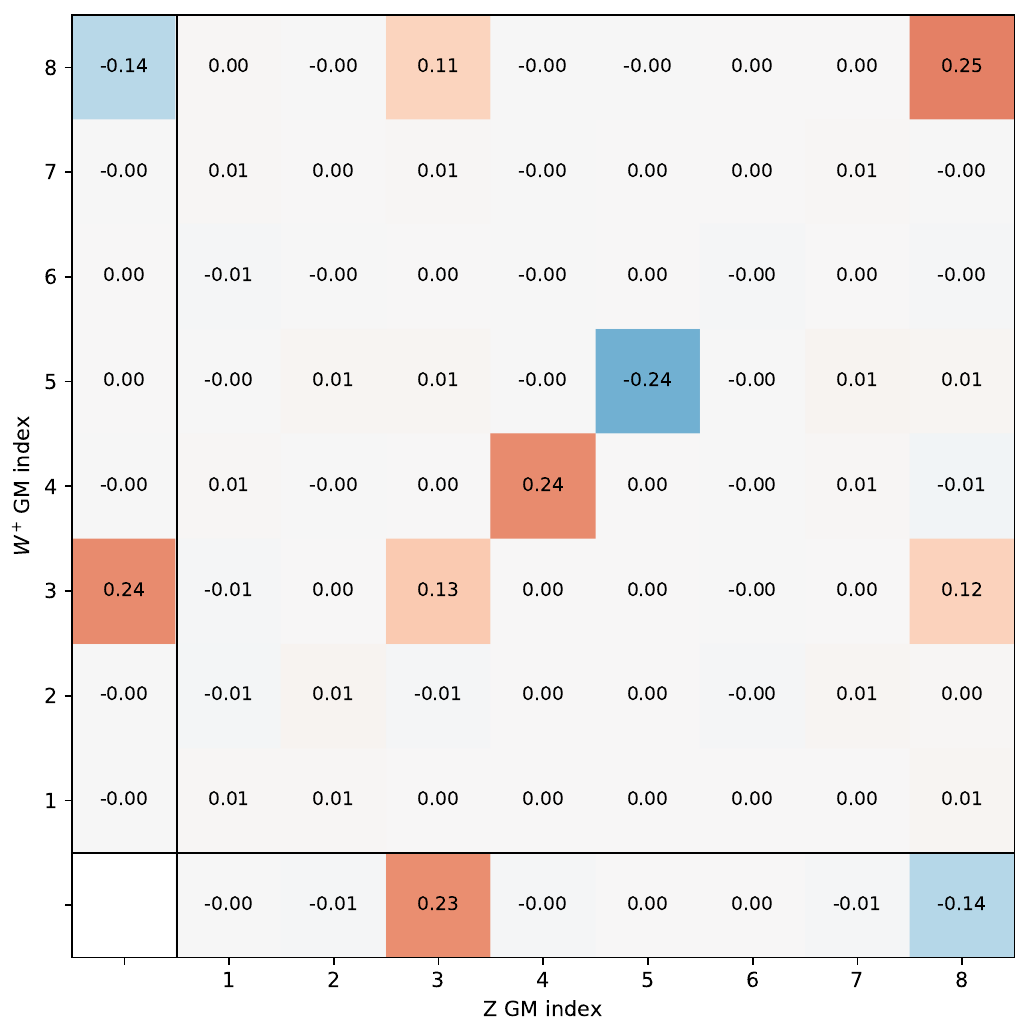}
  \caption{Gell-Mann coefficient matrix for the \(W^+Z\) system: even quadratic contribution.}
  \label{fig:wz_cij_even_quad}
\end{figure}

\begin{figure}[H]
  \centering
  \includegraphics[width=0.8\linewidth]{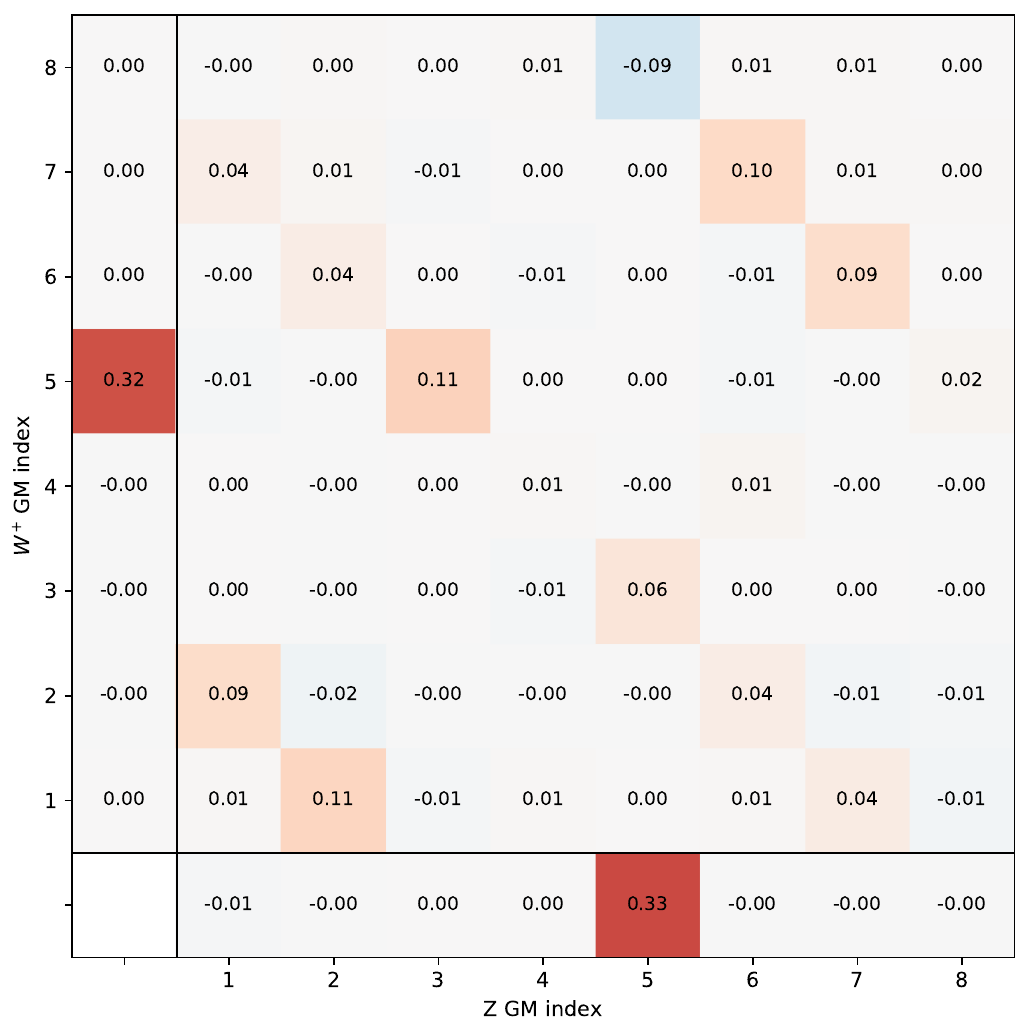}
  \caption{Gell-Mann coefficient matrix for the \(W^+Z\) system: odd interference contribution.}
  \label{fig:wz_cij_odd_int}
\end{figure}

\begin{figure}[H]
  \centering
  \includegraphics[width=0.8\linewidth]{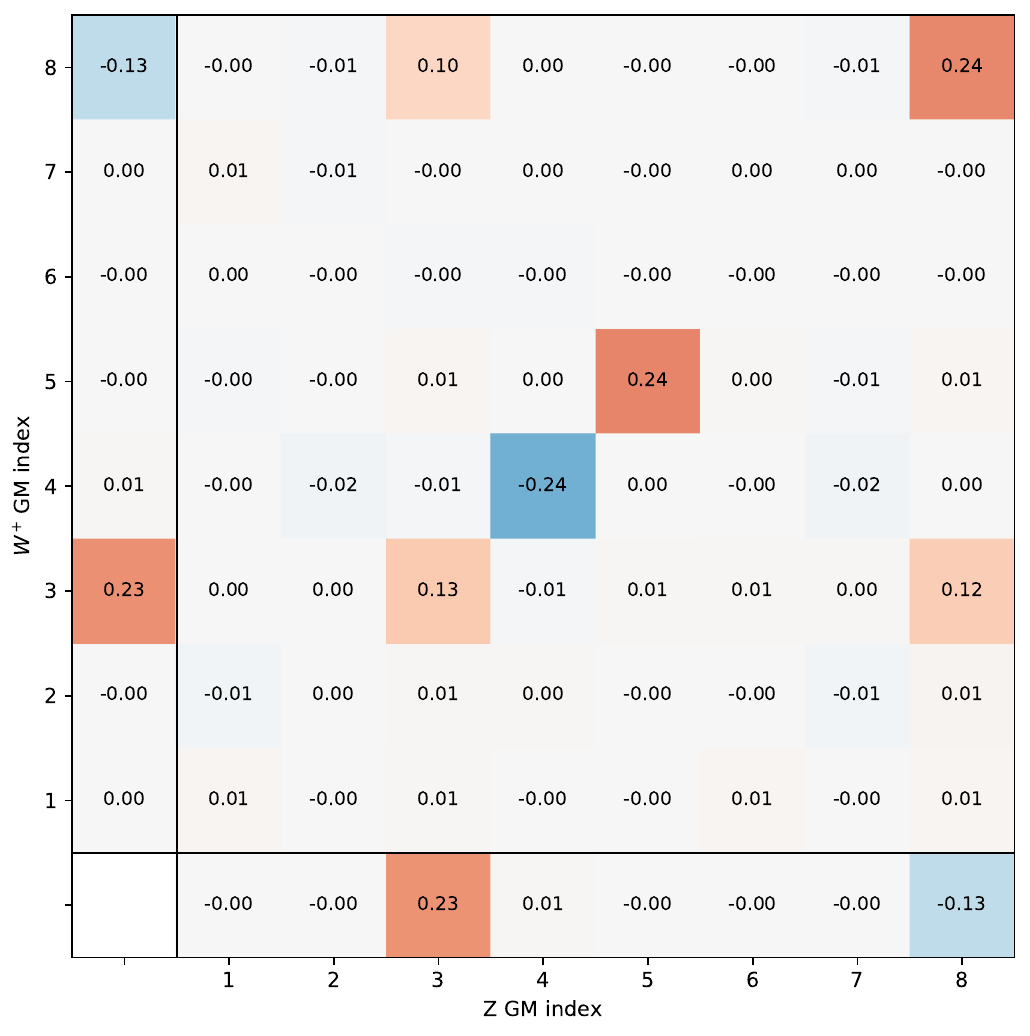}
  \caption{Gell-Mann coefficient matrix for the \(W^+Z\) system: odd quadratic contribution.}
  \label{fig:wz_cij_odd_quad}
\end{figure}

\printbibliography

\end{multicols}

\end{document}